\documentclass[%
aps,
prd,
twocolumn,
preprintnumbers,
nofootinbib,
amsmath,amssymb,
]{revtex4-2}

\usepackage{style}
\usepackage{aas_macros}

\begin{document}

%\preprint{ZU-XXXX/21}

\title{Back to the phase space: \\ thermal axion dark radiation via couplings to standard model fermions}

\author{Francesco D'Eramo}
\email{francesco.deramo@pd.infn.it}
\affiliation{Dipartimento di Fisica e Astronomia, Universit\`a degli Studi di Padova, Via Marzolo 8, 35131 Padova, Italy \\ 
Istituto Nazionale di Fisica Nucleare (INFN), Sezione di Padova, Via Marzolo 8, 35131 Padova, Italy}
\author{Alessandro Lenoci}
\email{alessandro.lenoci@mail.huji.ac.il}
\affiliation{Racah Institute of Physics, The Hebrew University, 91904, Jerusalem, Israel}
\begin{abstract}
We investigate the cosmological consequences of axion interactions with standard model fermions accurately and precisely. Our analysis is entirely based on a phase space framework that allows us to keep track of the axion distribution in momentum space across the entire expansion history. First, we consider flavor-diagonal couplings to charged leptons and quantify the expected amount of dark radiation as a function of the coupling strength. Leptophilic axions are immune from complications due to strong interactions and our predictions do not suffer from theoretical uncertainties. We then focus on flavor-diagonal interactions with the three heavier quarks whose masses are all above the scale where strong interactions become non-perturbative. The top quark case is rather safe because its mass is orders of magnitude above the confinement scale, and the consequent predictions are solid. The bottom and charm masses are in more dangerous territory because they are very close to the QCD crossover. We present a comprehensive discussion of theoretical uncertainties due to both the choice of the scale where we stop the Boltzmann evolution and the running of QCD parameters. Finally, we compute the predicted amount of dark radiation expressed as an effective number of additional neutrino species.  We compare our predictions with the ones obtained via standard approximate procedures, and we find that adopting a rigorous phase space framework alters the prediction by an amount larger than the sensitivity of future CMB observatories.
\end{abstract}

\maketitle
\allowdisplaybreaks
%\tableofcontents

%%%%%%%%%%%%%%%%%%%%%%%%%%%%%%%%%%%%%%%%%%%%%%%%
\section{Introduction}
\label{sec:intro}
%%%%%%%%%%%%%%%%%%%%%%%%%%%%%%%%%%%%%%%%%%%%%%%%

The lack of discoveries at high energy colliders and direct dark matter searches suggests that new physics may not arise at the Fermi scale after all. A rich and diverse experimental program has been exploring theories where hypothetical new particles are light but with interactions too small to be observed yet. Besides being experimentally accessible via this intensity frontier program, this landscape of theoretical frameworks contains several options motivated from the top down.

Pseudo-Nambu-Goldstone bosons (PNGB) are appealing as they are naturally light due to the shift symmetry that they inherit from UV dynamics. They are ubiquitous in motivated frameworks such as string compactifications and spontaneously broken global symmetry in quantum field theory. The Peccei-Quinn (PQ) framework~\cite{Peccei:1977np,Peccei:1977hh}, which was originally proposed to address the puzzle of CP conservations in Quantum ChromoDynamics (QCD), provides what is today one of the most appealing and searched for PNGB: the QCD axion~\cite{Wilczek:1977pj,Weinberg:1977ma}. Axion detection is rather challenging as a consequence of the extremely feeble couplings between this hypothetical new degree of freedom and visible particles. Search strategies include seeking to detect an axion dark matter abundance~\cite{Preskill:1982cy,Abbott:1982af,Dine:1982ah}, which can succeed only if a substantial cosmic abundance in the Solar system neighborhood is present, and other experimental techniques that do not rely upon the axion comprising the totality of dark matter such as effects of long-range axion-mediated forces. 

The early universe provides us with a unique laboratory to test axion dynamics. We know that the energy budget was dominated by relativistic components at the time of Big Bang Nucleosynthesis (BBN). If we extrapolate such a BBN snapshot a few orders of magnitude back in time and up to temperature at least to the weak scale, we find a primordial thermal bath populated by all standard model particles. This bath is an \textit{unavoidable} production source for relativistic axions that get dumped into the early universe and may or may not reach thermal equilibrium. Such an \textit{inevitability} has been powerful to put constraints in other contexts as well~\cite{Langhoff:2022bij,DEramo:2024lsk,Iles:2024zka}.

In this work, we focus on axion production in the early universe via binary scatterings mediated by axion flavor-conserving interactions with standard model fermions. First, we consider the cases for the three charged leptons where only electromagnetic interactions play a role since there are no issues associated with confinement and the QCD crossover does not pose any threat to the reliability of our calculations. Then we consider couplings to the charm, bottom, and top quarks, for which we start having QCD effects that make the analysis potentially more involved. The top quark mass is far from the confinement scale and in a region where QCD is perturbative. The situation is more concerning for the bottom quark, and it is especially alarming for the charm quark. We discuss the uncertainties associated with the QCD crossover being so close to the production epoch as well as the effects of QCD running parameters.

The novel feature of our analysis is that all predictions are entirely based on a phase space framework. In particular, we solve the Boltzmann equation in momentum space for the axion distribution function and we keep track of this quantity across the expansion history. It is hard to overemphasize the benefits of a formalism entirely based on a phase space analysis, and the associated advantages are multiple. First of all, the assumption about axions achieving thermal equilibrium at some point in the past is never necessary, and we can study light relics that were never coupled to the bath. For couplings large enough to ensure thermalization, a phase space analysis is sensitive to the decoupling epoch and it is capable of quantifying potential spectral distortions in the distribution that would lead to detectable effects. Both bath and dark radiation particles can be treated quantum mechanically in a self-consistent way. Energy conservation is ensured by accounting for the energy exchanged between dark and visible sectors. How this formalism can be implemented in a general particle physics framework as well as a comprehensive discussion of the associated benefits can be found in Ref.~\cite{DEramo:2023nzt}.

With the asymptotic axion phase space distribution in hand, we make connections with physical observables. In this regard, a central question is how this cosmic axion population would manifest itself once one analyzes cosmological data. They are relativistic at the production epoch, and they would eventually propagate along the geodesics of the expanding universe with energy decreasing inversely proportional to the scale factor. Regardless of whether they achieve thermal equilibrium or not, their typical energy would not be too different from the temperature of the thermal bath; this is reasonable given their thermal origin, and it is also something that we check explicitly every time in this paper by computing an effective axion temperature. At the BBN epoch, when the bath temperature is approximately around the MeV scale, axions are still relativistic and contribute to the radiation energy density. The amount of dark radiation is historically parameterized in terms of an effective number of additional neutrino species $\dneff$. We measure this quantity at the BBN time with exquisite precision, and the possible presence of dark radiation at that epoch is quite constrained~\cite{Yeh:2020mgl,Pisanti:2020efz,Yeh:2022heq}. Another key phase of the expansion history is at the time of the formation of the Cosmic Microwave Background (CMB) when the bath temperature was around the eV scale. We measure $\dneff$ also at this epoch~\cite{Planck:2018vyg,ACT:2020gnv,SPT-3G:2021wgf}, and if the axion mass is well below the eV scale then we can put a further constraint on the dark radiation energy density. Current constraints from BBN and CMB are competitive and lead to the upper bound $\dneff \lesssim 0.2$. In the near future, we are going to improve significantly the sensitivity on $\neff$ via the Simons Observatory~\cite{SimonsObservatory:2018koc} and CMB-S4~\cite{CMB-S4:2016ple,Abazajian:2019eic,CMB-S4:2022ght}. Futuristic proposals~\cite{Sehgal:2020yja} aim to reach $\sigma(\dneff) \simeq 0.01$. This will further limit the presence of dark radiation in the early universe, and it could open a window into the dark sector if a non-vanishing $\dneff$ is detected.

Thermal axion abundance calculations and the associated predictions for $\dneff$ have a long story. The first analysis was performed by Ref.~\cite{Turner:1986tb} almost 40 years ago. Other earlier studies on thermal axion production, but via pion scatterings, can be found in Refs.~\cite{Berezhiani:1992rk,Chang:1993gm}. Axion production via the coupling to gluons, which is always present if we are dealing with the QCD axion solving the strong CP problem, has been regularly improved and discussed over the last 20 years~\cite{Masso:2002np,Graf:2010tv,Salvio:2013iaa,DEramo:2021psx,Bouzoud:2024bom}. Axion couplings to standard model fermions, both flavor-diagonal and flavor-violating interactions, were analyzed within the instantaneous decoupling approximation~\cite{Baumann:2016wac} and via ordinary differential Boltzmann equations tracking the axion number density~\cite{Ferreira:2018vjj,DEramo:2018vss,Arias-Aragon:2020qtn,Arias-Aragon:2020shv,Green:2021hjh,DEramo:2021usm}. Ref.~\cite{DEramo:2021lgb} considered both axion couplings to bosons and fermions, and employed the Boltzmann equation for the axion number density to predict $\dneff$ for the two most popular UV complete frameworks: KSVZ~\cite{Kim:1979if,Shifman:1979if} and DFSZ~\cite{Zhitnitsky:1980tq,Dine:1981rt}. All these studies neglected the axion mass, or in other words, they assumed it to be well below the CMB temperature at the recombination epoch. Finite axion mass effects were explored by Refs.~\cite{Ferreira:2020bpb,Caloni:2022uya,DEramo:2022nvb,Notari:2022ffe,Bianchini:2023ubu}. This work will push this effort one step forward by studying the cosmological consequences of axion-fermion interactions via a rigorous momentum space analysis.  As we will show, spectral distortions that are generated due to a non-instantaneous decoupling alter the prediction for $\dneff$ with respect to the ones obtained via standard methods employing the Boltzmann equation for the axion number density. We find mismatches between predictions obtained via rigorous and approximate methods that can be larger than the experimental sensitivities of futuristic CMB surveys.

This paper is structured as follows. Sec.~\ref{sec:thermalization} studies axion (re)coupling at low temperatures, and we show with explicit examples how axion couplings to fermions could lead to IR thermalization. Sec.~\ref{sec:PS} briefly summarizes the phase space formalism introduced by Ref.~\cite{DEramo:2023nzt} and employed by this work. We divide the following discussion into two main parts corresponding to axion couplings to leptons and quarks, and this is contained in Secs.~\ref{sec:leptophilic} and \ref{sec:hadrophilic}, respectively. The material of these sections is rather technical. Readers interested in the details of the phase space analysis will find every step of our investigation as well as a discussion of the theoretical uncertainties. Our final predictions for $\dneff$ are provided in Sec.~\ref{sec:dneff} together with a comparison with the analogous ones obtained via approximate analysis. In particular, we compare our findings with the ones via the solution of the Boltzmann equation tracking the evolution of the axion number density. We collect the astrophysical bounds on the axion-fermion interactions studied here in App.~\ref{app:bounds}, and we review the approximate method based on the number density in App.~\ref{app:BE4n}. The main results are summarized in the conclusions given in Sec.~\ref{sec:conclusions}. In particular, the two panels of Fig.~\ref{fig:NeffFinal} show the predicted $\dneff$ as a function of the fermion couplings for each case and the comparison between cosmological and astrophysical bounds, respectively.

%%%%%%%%%%%%%%%%%%%%%%%%%%%%%%%%%%%%%%%%%%%%%%%%
\section{UV/IR Axion Thermalization}
\label{sec:thermalization}
%%%%%%%%%%%%%%%%%%%%%%%%%%%%%%%%%%%%%%%%%%%%%%%%

Axion interactions with particles belonging to the primordial plasma are responsible for thermal production at early times. Under the most conservative assumption of a thermal bath populated only by standard model particles, we classify axion interactions as follows
\be
\mathcal{L}_{\rm int} = \mathcal{L}^V_{\rm int} + \mathcal{L}^\psi_{\rm int} \ .
\label{eq:Lint}
\ee 
The two terms in the Lagrangian contain axion couplings to gauge bosons $V$ and fermions $\psi$, respectively. This paper investigates the cosmological consequences of the latter only. However, it is beneficial to discuss axion thermalization generally to put our specific analysis into a broader perspective. In this section, we account for all possible interactions in Eq.~\eqref{eq:Lint} and examine at what temperatures and for what coupling strengths it is possible to achieve thermal equilibrium. In particular, we identify whether a given process is IR or UV-dominated.

Interactions with gauge bosons arise whether the spontaneously broken global symmetry (e.g., the PQ symmetry for the QCD axion) is anomalous under that specific gauge group. At high energies, well above the weak scale, the axion $a$ can couple potentially to all standard model gauge bosons and these couplings respect the electroweak $SU(2)_L \times U(1)_Y$ gauge invariance. Once one reaches temperatures below the weak scale, the only active interactions are the ones with QCD and QED gauge bosons 
\be
\mathcal{L}^V_{\rm int} = \frac{\ax}{8 \pi f_\ax} \left( \alpha_s \widetilde{G}^{\mu \nu} G_{\mu \nu} + c_\gamma \, \alpha_{\rm em} \widetilde{F}^{\mu \nu} F_{\mu \nu} \right)\ .
\label{eq:LintV}
\ee
We normalize them as it is conventionally done in the literature. The coupling to the gluon field strength $G_{\mu\nu}$ (and its dual $\widetilde{G}^{\mu\nu} \equiv \epsilon^{\mu\nu\rho\sigma} G_{\rho\sigma}$) is proportional to the fine structure constant $\alpha_s \equiv g_s^2 / (4 \pi)$ of the strong gauge group, and all adjoint indices are implicitly understood and summed over. There is no further dimensionless Wilson coefficient and therefore we take this operator as the definition of the axion decay constant $f_\ax$. On the contrary, the interaction with the electromagnetic field strengths has the same Lorentz structure and normalization in terms of the fine structure constants $\alpha_{\rm em} \equiv e^2 / (4 \pi)$, but it also has a model-dependent Wilson coefficient $c_\gamma$. These operators mediate UV-dominated production since they have mass dimension 5 and there are no other dimensionful quantities that can appear in the rate besides the axion decay constant suppressing the rate. In particular, the number of interactions per unit time $\Gamma_V$ due to the interaction with the gauge boson $V$ scales as $\Gamma_V \propto c_V^2 \alpha_V^3 T^3 / f_\ax^2$ (for gluons we have $c_G = 1$). Thus these interactions are always capable of bringing the axion to thermal equilibrium as long as the reheat temperature after inflation is large enough. Thermalization at high temperatures does not lead to a strong observable signal since the contribution to $\dneff$ is inversely proportional to a positive power of the effective number of degrees of freedom at the decoupling temperature $T_{\rm d}$ ($\dneff \propto g_{*s}(T_{\rm d})^{-4/3}$). As is well known, light (pseudo-)scalars that decoupled above the weak scale are barely within reach of future CMB surveys. This makes interactions with gauge bosons possibly less appealing.\footnote{What written in this paragraph is true as long as the dimension 5 operators in Eq.~\eqref{eq:LintV} remain local at the temperatures of interest. For example, the operator with gluons in the KSVZ framework is not local at temperatures above the mass of the heavy and colored PQ-charged fermion that is responsible for generating that coupling. At higher temperatures, axion production becomes IR-dominated. See Ref.~\cite{DEramo:2021lgb} for a detailed discussion.} 

The focus of the present work is on axion interactions with standard model fermions. Consistently with the Nambu-Goldstone nature of the axion field $\ax$, the lowest dimensional couplings of this kind are between the axion spacetime derivative $\partial_\mu \ax$ and spin-one fermion currents. Each operator of this class is suppressed by the axion decay constant $f_\ax$, and it is multiplied by a dimensionless model-dependent Wilson coefficient. Remarkably, these interactions mediate \textit{both} UV and IR-dominated production. At high temperatures, before the breaking of the electroweak symmetry and therefore when all standard model fermions are massless, the only possible processes are binary scatterings involving the four components of the standard model Higgs doublet. Processes with gauge bosons on the external legs have vanishing amplitudes because there is a fermion chirality flip needed in the amplitude and fermion masses vanish in this phase~\cite{Arias-Aragon:2020shv}. Thus the axion production rate at high temperatures in this case scales as $\Gamma_\psi \propto y_\psi^2 T^3 / f_\ax^2$, where $y_\psi$ is the fermion Yukawa coupling, and the production is still UV dominated. Thermalization is more efficient for heavy fermions and it is always achieved for a large enough reheat temperature. The conclusions are analogous to the ones in the previous paragraph.

\begin{figure*}[!t]
	\centering
		\includegraphics[width=0.49\textwidth]{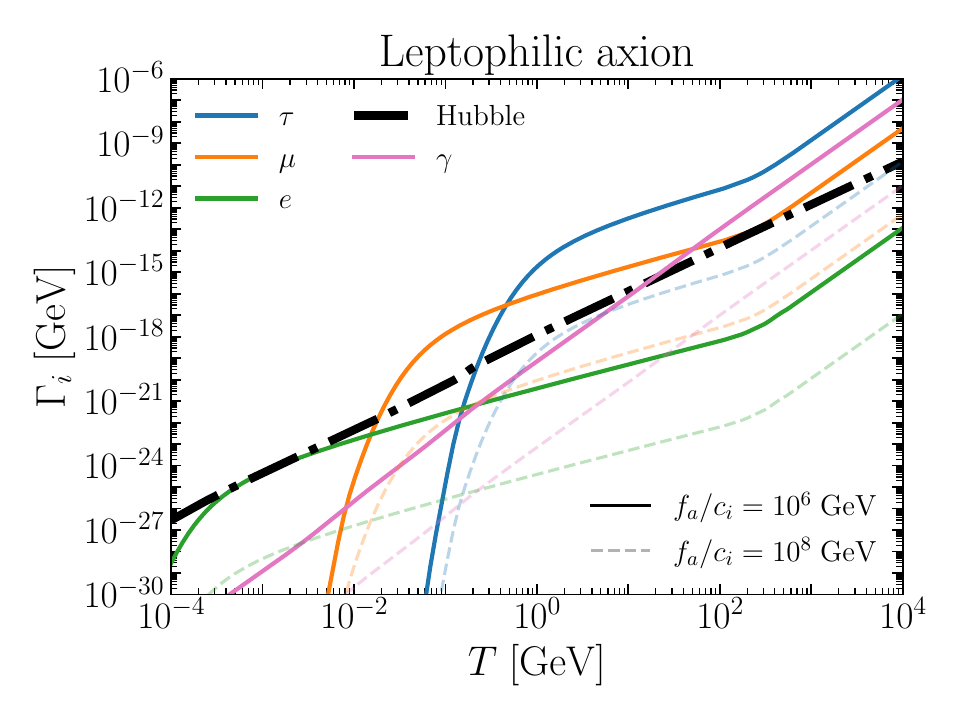} $\;$
	\includegraphics[width=0.49\textwidth]{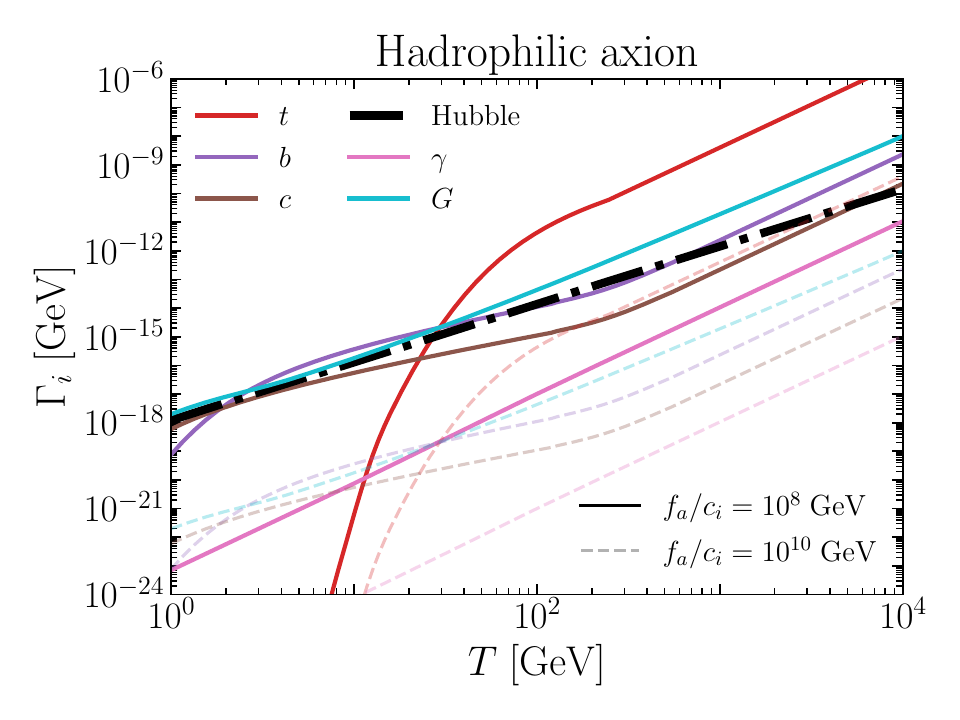}
	\caption{Axion production rates as a function of the temperature for leptophilic axion (left panel) and hadrophilic axion (right panel). Colored lines, both solid and dashed, show the production rates when only the interaction with that specific standard model particle is switched on in the Lagrangian. The difference between solid and dashed lines is the chosen value for $f_\ax / c_i$ as explained in the legenda. The dashed-dot black line shows the temperature-dependent Hubble expansion rate.}
	\label{fig:rates}
\end{figure*}

The situation is drastically different at temperatures below the weak scale. As we will show soon, these interactions mediate an IR-dominated production since the collision rate is more and more efficient as we approach the mass of the fermion itself. With the exception of the top quark, all the other fermions have a mass well below the weak scale. The top quark case is borderline, but if axions thermalize then interactions are effective even at temperatures slightly below the top mass when the Boltzmann suppression starts to kick in. The work in Ref.~\cite{Arias-Aragon:2020shv} provided calculations for the axion production mediated by the top coupling above and below the weak scale and showed a smooth connection between the two regimes across the electroweak phase transition.

We consider scenarios where axion couplings are flavor-diagonal in the mass eigenbasis
\be
\mathcal{L}^\psi_{\rm int} = \frac{\partial_\mu \ax}{2f_\ax} \sum_\psi c_\psi \overline{\psi} \gamma^\mu \gamma^5 \psi  \ .
\label{eq:LintPsi}
\ee
The Dirac field appearing in the sum are charged leptons ($\psi = \{e, \mu, \tau \}$), and heavy quarks ($\psi = \{c, b, t \}$).  As anticipated above, the production becomes IR-dominated at temperatures below the weak scale, and employing Dirac fermions in this phase is convenient. We do not include operators proportional to vector currents since they vanish upon integrating by parts (up to electroweak anomalous terms irrelevant to our analysis). In the broken phase, axions are produced via fermion scatterings involving one gauge boson $V$ in the initial or final states, and the production rate scales as $\Gamma_\psi \propto \alpha_V c_\psi^2 m_\psi^2 T / f_\ax^2$ as long as the temperature $T$ is above the fermion mass. The factor of $m_\psi^2$ is the consequence of the aforementioned fermion chirality flips. Thus the rate scales with a weaker temperature power than the Hubble rate, and the axion production rate is most efficient at temperatures around the fermion mass. 

The cosmological consequences of the interactions in Eq.~\eqref{eq:LintPsi} are the main subject of this work. The reason why this section also contains the interactions with gauge bosons, as given in Eqs.~\eqref{eq:LintV} below the weak scale, is because we find it useful to identify the parameter space regions where fermion couplings are indeed responsible for axion production and thermalization. One straightforward observation is that the axion decay constant $f_\ax$ suppresses all the mass-dimension 5 interactions, and the relative sizes of the Wilson coefficients $c_i$ play a crucial role. Here, we assume no specific hierarchy among these dimensionless couplings and we take $c_i \simeq 1$ when we do the comparison.

Fig.~\ref{fig:rates} shows such a comparison explicitly for axion interactions with leptons (left panel) and heavy quarks (right panel). Both panels illustrate the temperature-dependent axion production rates for interactions with different particles (corresponding to different colors) and for two different choices for $f_\ax / c_i$ (solid and dashes). Each panel reports also the Hubble expansion rate $H(T)$ as a function of the temperature. Axions achieve thermal equilibrium when $\Gamma_i \gtrsim H$, and maintain this condition as long as such an inequality is satisfied.

The left panel of Fig.~\ref{fig:rates} includes axion coupling to the three charged leptons and photons. We notice how the photon coupling has a hard time bringing axions to thermal equilibrium, and it is nevertheless a subdominant contribution with respect to fermion couplings even when it can do so. The two choices for $f_\ax / c_i$ show two rather different scenarios. If the interactions are feeble enough, as is the case for the dashed lines, thermalization is never achieved. The coupling strengths are increased by two orders of magnitude in the rate shown by the solid lines, and thermalization is achieved at high temperatures. Furthermore, we note a peculiar behavior of the colored lines as a consequence of the electroweak phase transition. The muon coupling is brought almost out of equilibrium around the weak scale, but then then recovers quickly and wins over Hubble. The electron coupling is an even more extreme case: it is never in thermal equilibrium at high temperatures, and it thermalizes much later in the expansion history. Thus having solely IR thermalization is possible. The discussion for the right panel of Fig.~\ref{fig:rates} is analogous. Even in that case, it is possible to have particles that are coupled at very high temperatures in the UV that then decouple and recouple in the IR below the weak scale. 

To summarize, UV thermalization above the weak scale is possible at the price of a rather large reheat temperature. Furthermore, the observable signal is only accessible to future CMB surveys by the narrowest of margins. This makes IR thermalization more appealing for detecting the observable effects. Axion couplings to fermions make this possible, and we will study it in the next sections via a rigorous phase space analysis. 

%%%%%%%%%%%%%%%%%%%%%%%%%%%%%%%%%%%%%%%%%%%%%%%%
\section{Back to the phase space}
\label{sec:PS}
%%%%%%%%%%%%%%%%%%%%%%%%%%%%%%%%%%%%%%%%%%%%%%%%

We summarize briefly the phase space formalism that allows us to track the momentum distribution of axions throughout cosmic history. Readers interested in the details can find plenty in Ref.~\cite{DEramo:2023nzt}. The next two sections contain applications to leptophilic and hadrophilic axions, respectively. The axion phase distribution does not depend on spatial coordinates (homogeneity) and the direction of the momentum (isotropy). Thus it can only depend on the size of the spatial momentum $k$ and the cosmic time $t$. We define $\f_\ax(k,t)$ the axion phase space distribution. The axion mass is always assumed to be much smaller than the bath temperature at the last scattering surface and therefore negligible. Given the couplings in Eq.~\eqref{eq:LintPsi}, we have to account only for the production process with one axion particle in the final state. These are the dominant production channels because every time we make an axion there is a price to pay due to the feeble axion couplings suppressed by the large scale $f_\ax$. The Boltzmann equation valid for such a \textit{single production} scenario is rather simple
\be
\frac{d\f_\ax(k, t)}{dt} = \left( 1 - \frac{\f_\ax(k, t)}{\f_\ax^{\rm eq}(k, t)}  \right)  \mathcal{C}_\ax (k, t) \ .
\label{eq:Cfsingle1}
\ee
Here, $\f_\ax^{\rm eq}(k, t)$ is the axion equilibrium (Bose-Einstein) distribution. All dynamics are encoded in the collision term ${\cal C}_\ax$ which obtains contributions from multiple processes, ${\cal C}_\ax  = \sum_\mathcal{P} {\cal C}_\mathcal{P}$. For a generic scattering $\bath_i \bath_j \to \bath_\ell \ax$, where $\bath_i$ are generic particles belonging to the primordial bath, the collision term results in
\bea\nonumber
{\cal C}_{\bath_i \bath_j \to \bath_\ell \ax}(k,T) &=& \dfrac{1}{2\omega}\int d\Pi_id\Pi_j d\Pi_\ell\\\nonumber
&&  \times (2\pi)^4 \delta^{(4)}(P_i + P_j  -P_\ell -K)\\
&&  \times |{{\cal M}_{\bath_i \bath_j \to \bath_\ell \ax}|^2 \f_{i}\f_{j}(1\pm \f_{\ell})}\ . 
\label{eq:collterm}
\eea 
The axion energy is equal to the size of its physical momentum, $\omega = k$, because we neglect the axion mass. Consistently with the convention established by Ref.~\cite{DEramo:2023nzt}, the Lorentz invariant phase space factors for all bath particles $\bath_i$ contain the multiplicities of the internal degrees of freedom and the squared matrix element is averaged over both initial and final states. Finally, $\f_i$ is the phase space distribution of the $\bath_i$ particle. Quantum effects for bath particles are manifestly accounted for while for the axion they are taken into account by $\f_\ax^{\rm eq}$ in Eq.~\eqref{eq:Cfsingle1}. 

Axion production processes subtract energy from the thermal bath. We couple Eq.~\eqref{eq:Cfsingle1} with the Boltzmann equation accounting for the bath evolution to ensure that energy is conserved. There is no need to work in momentum space in this case since standard model interactions lead to processes in the thermal bath sector with rates that significantly exceed the expansion rate. Thus bath particles are always in local thermal equilibrium and we can work with a Boltzmann equation tracking the total bath energy density
\be
\begin{split}
& \frac{d \rho_\bath}{dt} + 3 H (1 + w_\bath) \rho_\bath = \\ &  - g_\ax \int \frac{d^3 k}{(2 \pi)^3}
k \, \left( 1 - \frac{\f_\ax(k, t)}{\f_\ax^{\rm eq}(k, t)}  \right)  \mathcal{C}_\ax (k, t) \ ,
\end{split}
\label{eq:BEbath}
\ee
with $g_\ax = 1$ and $w_\bath$ the equation of state parameter~\cite{Laine:2015kra}. The Hubble rate is given by the Friedmann equation
\be
H = \frac{1}{\sqrt{3} M_{\rm Pl}} \sqrt{\rho_\bath + \rho_\ax} \ ,
\label{eq:Friedmann}
\ee
where $M_{\rm Pl} \equiv (8 \pi G)^{-1/2} \simeq 2.4 \times 10^{18} \, {\rm GeV}$ is the reduced Planck mass.

\subsection{Numerical strategy}

Our goal is to solve the system of the three differential equations given explicitly by Eqs.~\eqref{eq:Cfsingle1}, \eqref{eq:BEbath} and \eqref{eq:Friedmann}. The variables appearing in them are not the most suitable for our purposes. For a numerical analysis, it is convenient to switch to dimensionless and comoving quantities. 

The scale factor $a(t)$ is a convenient evolution variable, and we trade the cosmic time $t$ with it. We solve the Boltzmann system starting from some initial time $t_I$ corresponding to $a_I = a(t_I)$. We define $T_I$ as the bath temperature at the time $t_I$; our results are not sensitive to the specific value for this variable if we choose it large enough since we are investigating IR thermalization. Furthermore, we introduce the ratio $A \equiv a / a_I$ and employ it as our evolution variable.

Physical momenta are not advantageous either. The Hubble expansion is responsible for their red-shift, and even after axion decoupling they decrease as $k \propto a^{-1}$ until the present time. This is why we employ the comoving momenta $q \equiv k A / T_I$. For the same reason, we work with comoving energy densities $R_i \equiv \rho_i A^4 / T_I^4$ for both the thermal bath ($i=\bath$) and the axion $(i = \ax)$.

We rewrite the Boltzmann system in terms of these new variables and integrate it from $A=1$, which is taken to correspond to an initial temperature $T_I=10^3 \, m_\psi$, to the final $A_F$ such that the corresponding temperature is well below the electron mass, $T_F\ll m_e$. The output of this numerical procedure is snapshots of the axion momentum distribution at any time of the cosmic history.

\subsection{Axion temperature}

One of the strengths of the phase space formalism is that it allows us to compute the full axion momentum distribution. This is not something directly observable, and in the end, we integrate to predict the value of $\dneff$. Another meaningful integrated quantity is the axion temperature defined from the width of the distribution~\cite{DEramo:2023nzt}
\be 
T_\ax(A) = \frac{T_I}{A}\bigg(\frac{\zeta(3)}{12 \, \zeta(5)}\bigg)^{1/2} \bigg[ \frac{\int dq \ q^4 \f_\ax(q,A)}{\int dq \ q^2\f_\ax(q,A)}\bigg]^{1/2}\ .
\label{eq:Ta}
\ee 
Here, $\zeta(x)$ is the Riemann function for the argument $x$, and the normalization factor ensures that we recover the temperature of a Bose-Einstein distribution when the interactions are strong enough to ensure thermalization. 

The definition in Eq.~\eqref{eq:Ta} is not the only one that reproduces the temperature of a Bose-Einstein. An analogous definition with the first moment of the distribution instead of the second would recover the equilibrium result upon choosing the normalization appropriately. However, for the most general distribution, the two results would differ. The scenario analyzed here, when equilibrium is not achieved, is analogous to dark matter freeze-in for which explicit distributions were derived in Ref.~\cite{DEramo:2020gpr}. It was also shown that the temperatures defined via the first and second moments are quite similar. The only observable effect is $\dneff$, the temperature is not observable no matter what definition we choose, we define a temperature is to have an estimate of the width of the distribution when we are far from equilibrium.  In what follows, we will adopt the definition in Eq.~\eqref{eq:Ta} and show how the axion temperature evolves with time.

\subsection{Determination of $\dneff$}

The observable consequence of a hot axion population is an additional contribution to the energy density stored in relativistic degrees of freedom quantified by an effective number of additional neutrino species 
\be\label{eq:dneff}
\dneff = \frac{8}{7} \left( \frac{11}{4} \right)^{4/3} \frac{\rho_\ax(T_{\rm CMB})}{\rho_\gamma(T_{\rm CMB})}  \ .
\ee
We do not need to solve the Boltzmann system until the recombination temperature $T_{\rm CMB}$, we can safely stop when the ratio $\rho_\ax / \rho_\gamma$ stays constant.
As stated above, we stop the Boltzmann evolution when the bath temperature $T_F$ is well below the electron mass, $T_F\ll m_e$. For such low temperatures, both axions and photons are decoupled and just red-shift as a relativistic fluid. Thus we can evaluate $\dneff$ as follows
\be
\dneff = \frac{8}{7} \left( \frac{11}{4} \right)^{4/3} \frac{g_{*\rho}(T_F)}{2} \frac{R_\ax(A_F)}{R_\bath(A_F)} \ .
\label{eq:dneffinal}
\ee

We also estimate the (numerical) error 
 \bea 
&& \delta \dneff = \frac{8}{7}\left( \frac{11}{4} \right)^{4/3}\frac{1}{2 R_\bath(A_F)/g_{*\rho}(T_F)} \\\nonumber
 &&\times \int_{\log A_I}^{\log A_F}d\log A \bigg|\frac{dR_\ax}{d\log A}+\frac{dR_\bath}{d\log A} + (3w_\bath -1 ) R_\bath \bigg| \ ,
 \eea 
where the second line estimates the contribution of the energy density that is numerically lost (the integrand should be exactly zero for energy conservation).  We check that in every solution the error is less than $10^{-3}$, with worst performances happening in the strongest coupling cases (i.e. where $\dneff$ is also the largest). In most of cases the error is below $10^{-4}$, but it can be further reduced by increasing the number of momentum bins.

%%%%%%%%%%%%%%%%%%%%%%%%%%%%%%%%%%%%%%%%%%%%%%%% 
\section{Leptophilic Axions}
\label{sec:leptophilic}
%%%%%%%%%%%%%%%%%%%%%%%%%%%%%%%%%%%%%%%%%%%%%%%%

Describing the phase space evolution for leptophilic axions is numerically challenging, but there are no further complications from particle physics. Charged leptons interact via electromagnetism only, gauge interactions are perturbative, and there are no other obstacles due to confinement. There are two main classes of production channels: lepton/antilepton pair annihilations, $\ell^+ \ell^- \to \gamma \ax$, and Compton-like scatterings for both leptons and antileptons, $\ell^{\pm} \gamma \to \ell^{\pm} \ax$. The squared matrix elements averaged above both initial and final states explicitly read
\bea
|{\cal M}_{\ell^+ \ell^- \to \ax \gamma }|^2 &=& \frac{c_\ell^2 e^2}{2f_\ax^2} \frac{m_\ell^2 s^2}{(m_\ell^2 - t)(s+t - m_\ell^2)} \ ,\\
|{\cal M}_{\ell^\pm \gamma \to \ell^\pm \ax }|^2 &=& \frac{c_\ell^2 e^2}{2f_\ax^2} \frac{m_\ell^2 t^2}{(s-m_\ell^2)(s+t - m_\ell^2)} \ ,
\eea 
where $e$ is the electron charge. The collision terms can be computed via Eq.~\eqref{eq:collterm} with the matrix elements above.

\begin{figure*}[!t]
	\centering
	\includegraphics[width=0.4\textwidth]{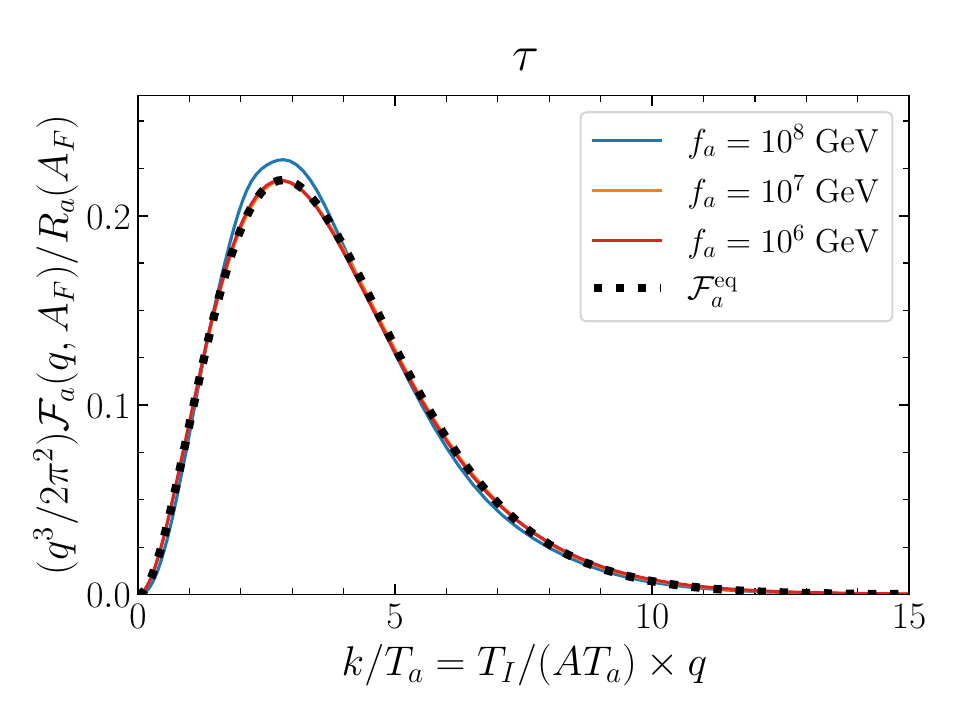} $\qquad$
	\includegraphics[width=0.4\textwidth]{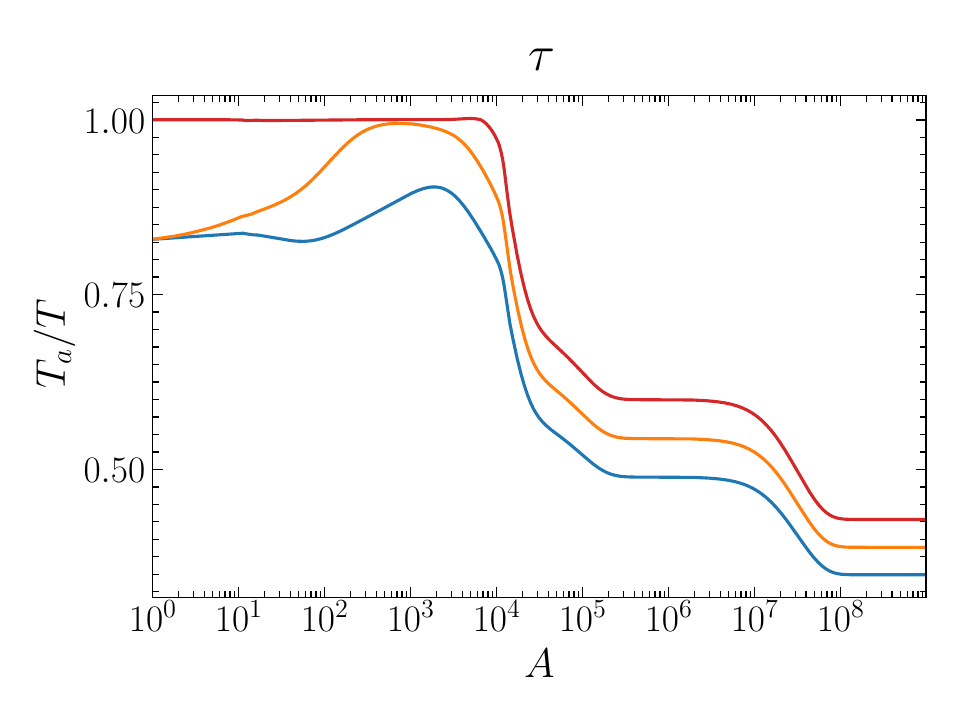}
	\\
		\includegraphics[width=0.4\textwidth]{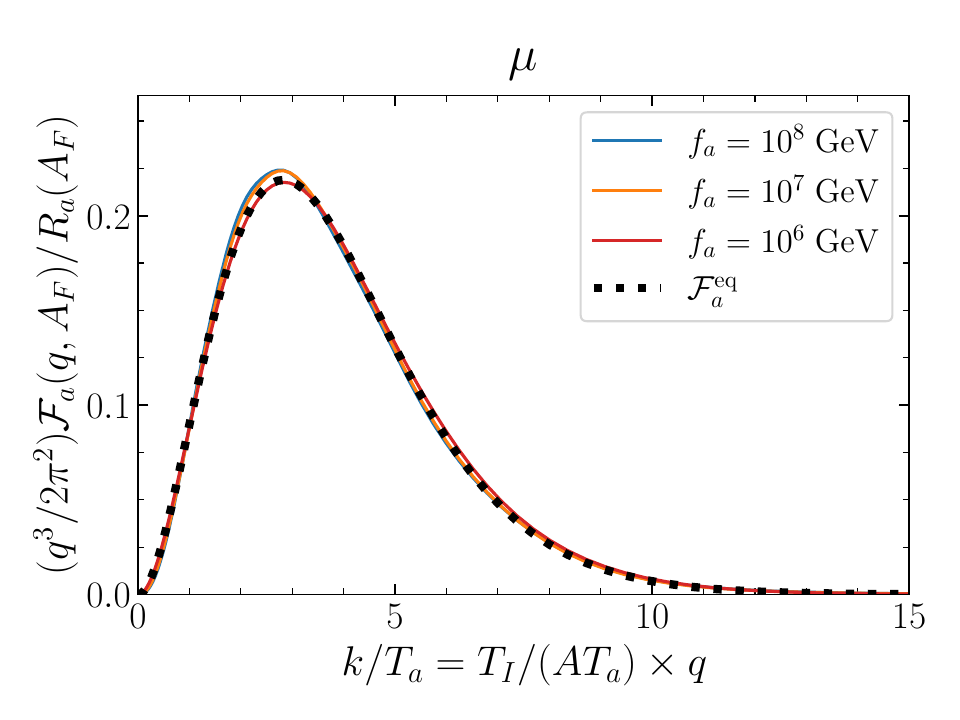} $\qquad$
	\includegraphics[width=0.4\textwidth]{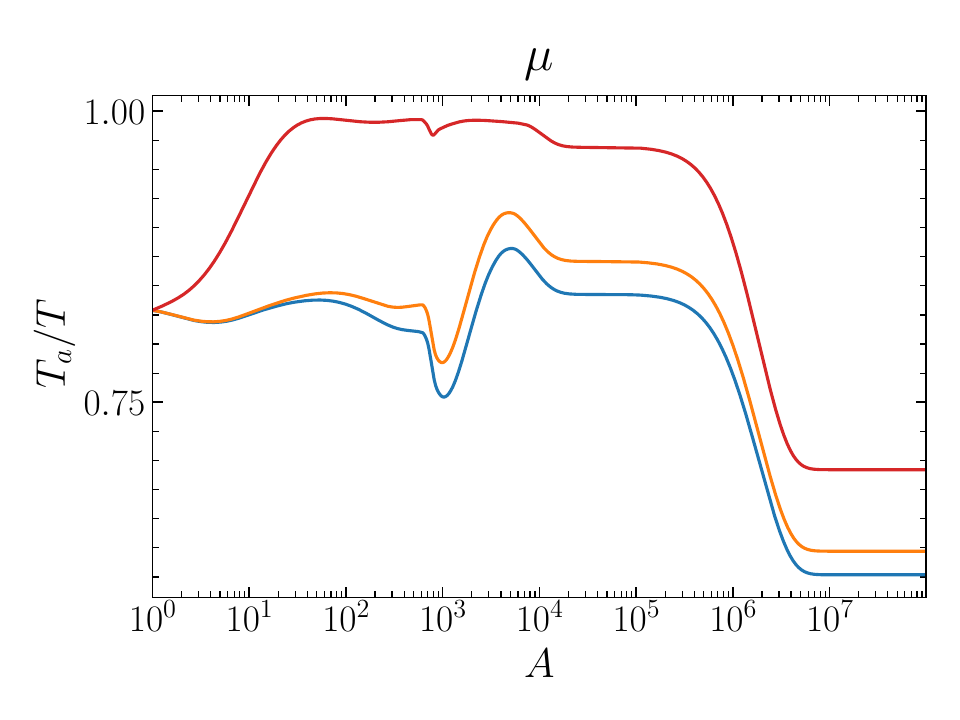}
	\\
	\includegraphics[width=0.4\textwidth]{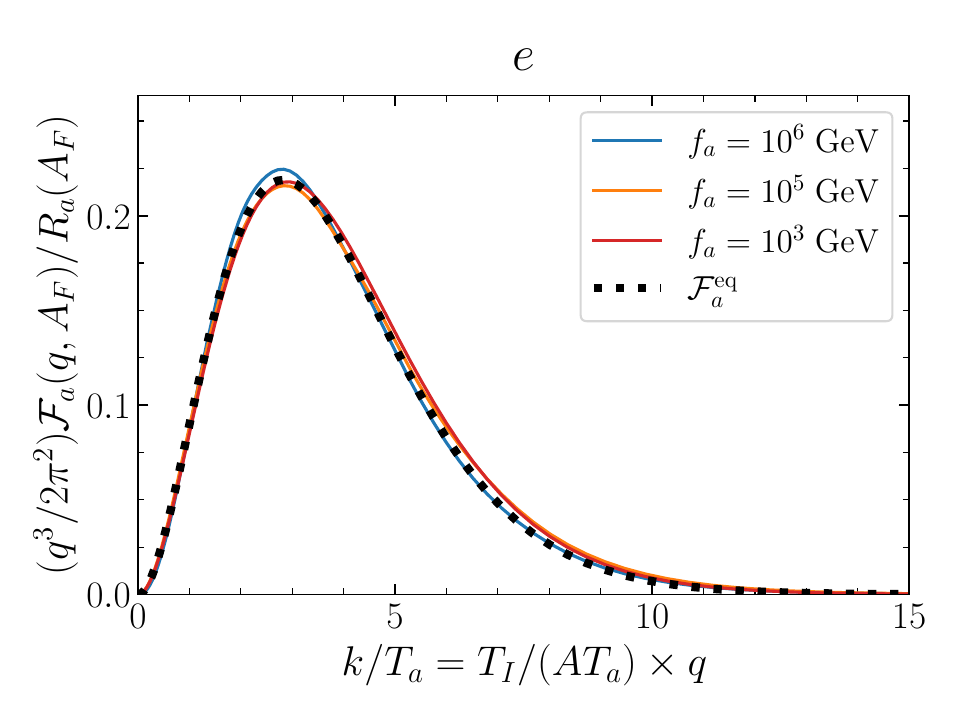} $\qquad$
	\includegraphics[width=0.4\textwidth]{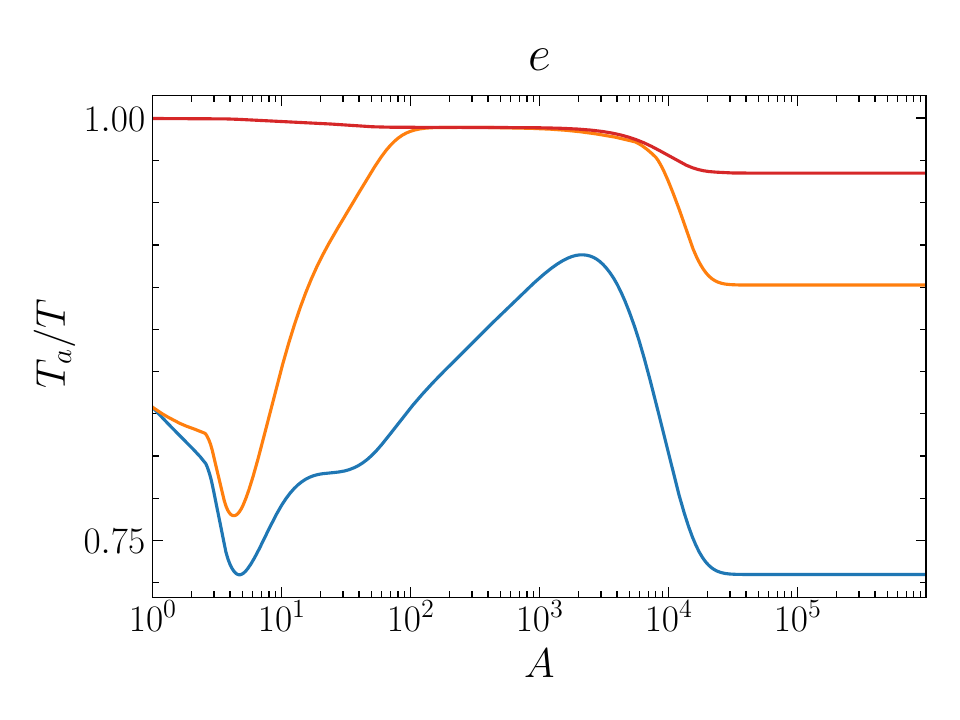}
\caption{Numerical results for leptophilic axions: asymptotic phase space distributions shown in terms of comoving variables defined in the main text (left) and evolution of the ratio $T_\ax / T$ between axion and bath temperatures (right). Different rows contain results for tau (top), muon (middle), and electron (bottom). Colored lines correspond to different choices for $f_\ax$ (all dimensionless Wilson coefficients are set to $c_\psi = 1$), and thick dotted lines identify the Bose-Einstein equilibrium distribution.}
\label{fig:PSDleptons}
 \end{figure*}

Fig.~\ref{fig:PSDleptons} summarizes the results of our numerical analysis. Each row contains results for a different charged lepton, and we present for each case both the momentum space distribution (left panels) and the axion temperature $T_\ax$ evolution (right panels). The choices for the variables on both axes of the left panels, which follow the conventions set by Ref.~\cite{DEramo:2023nzt}, deserve a few more comments. The asymptotic axion distribution $\f_\ax(q,A_F)$, expressed as a function of the comoving momentum $q$ and evaluated at the last step $A_F$ of the evolution, appears on the vertical axis multiplied by the phase space factor $q^3 / (2 \pi)^2$. The integral of this quantity gives the asymptotic comoving axion energy density expressed as defined above
\be
R_\ax(A_F) = g_\ax \int \frac{dq}{2 \pi^2} \, q^3 \f_\ax(q,A_F) \ .
\label{eq:Rax}
\ee
The resulting $\dneff$ is proportional to this quantity as expressed by Eq.~\eqref{eq:dneffinal}, and this is why we normalize the vertical axis by the same factor $R_\ax(A_F)$. Our goal here is to investigate deviations from thermal equilibrium and identify spectral distortions, and not to compare the relative sizes of $\dneff$. This is why we do not choose $q$ as the variable on the horizontal axis but the combination $T_I / (A T_\ax) \times q$. The rationale for this choice is understood upon recalling our definition of the comoving momentum which allows us to rewrite this variable as follows
\be
\frac{T_I}{A T_\ax} q =  \frac{k}{T_\ax} \ ,
\ee
where $k$ is the physical momentum. Thus this is the natural variable to employ since the equilibrium Bose-Einstein distribution reads $( \exp[k/ T_\ax] - 1)^{-1}$. We use the axion temperature $T_\ax$ defined in Eq.~\eqref{eq:Ta} rather than the bath temperature $T$ because we are focusing on asymptotic distributions evaluated at temperatures $T_F$ ($\ll m_e$) where axions do not need to share their temperature with the bath. Even for extreme case where axions have a perfect thermal distribution without any spectral distortions, they could have decoupled instantaneously earlier and the change of the number of relativistic degrees of freedom in the bath could have rescaled the relative temperatures. This is exactly what happens between neutrinos and photons in the standard cosmological model. 

If one chooses this clever set of variables then it is possible to investigate compactly in a single plot all deviations from thermal shapes for different choices of the coupling strengths. This is what the left panels of Fig.~\ref{fig:PSDleptons} illustrate. Axions produced from scatterings have a momentum distribution very close to a thermal shape identified by the black dotted line. This is true for large couplings where thermalization is achieved and for small couplings where axions do not get even close to equilibrium. The fact that the asymptotic distribution is always rather thermal up to an overall constant (the factor $R_\ax(A_F)$ in the normalization) makes sense given the thermal origin. Crucially, the resulting temperatures are different as shown by the right panels of Fig.~\ref{fig:PSDleptons}. Larger values of $f_\ax$ lead to smaller temperatures, and the net effect is a reduced contribution to $\dneff$ that we evaluate in Sec.~\ref{sec:dneff}.
 
%%%%%%%%%%%%%%%%%%%%%%%%%%%%%%%%%%%%%%%%%%%%%%%% 
\section{Hadrophilic Axions}
\label{sec:hadrophilic}
%%%%%%%%%%%%%%%%%%%%%%%%%%%%%%%%%%%%%%%%%%%%%%%% 

\begin{figure*}[!t]
	\centering
	\includegraphics[width=0.4\textwidth]{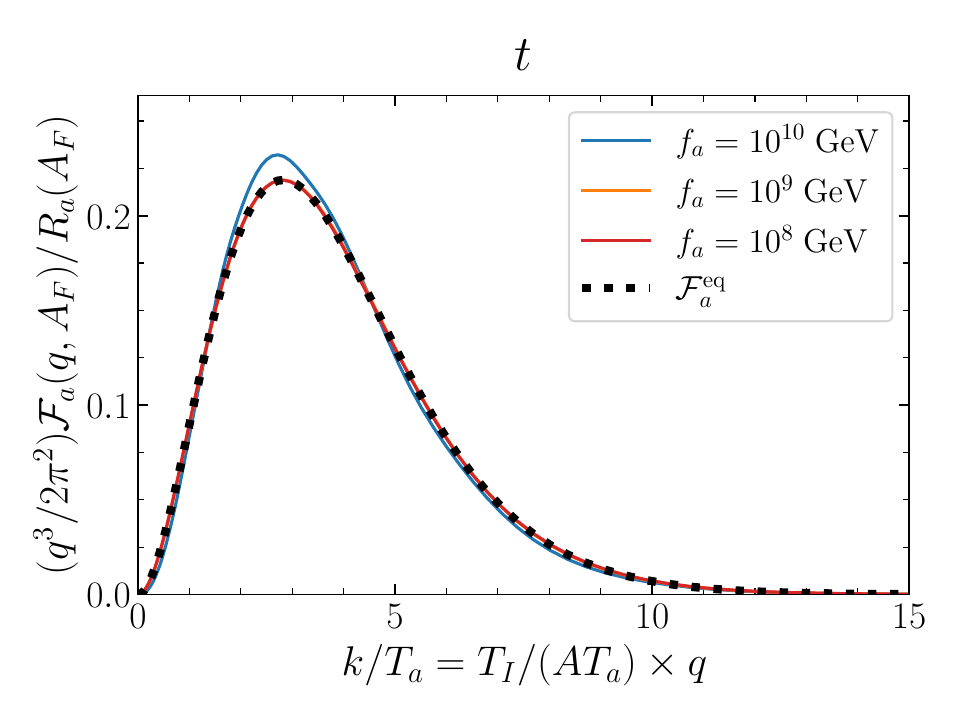} $\qquad$
	\includegraphics[width=0.4\textwidth]{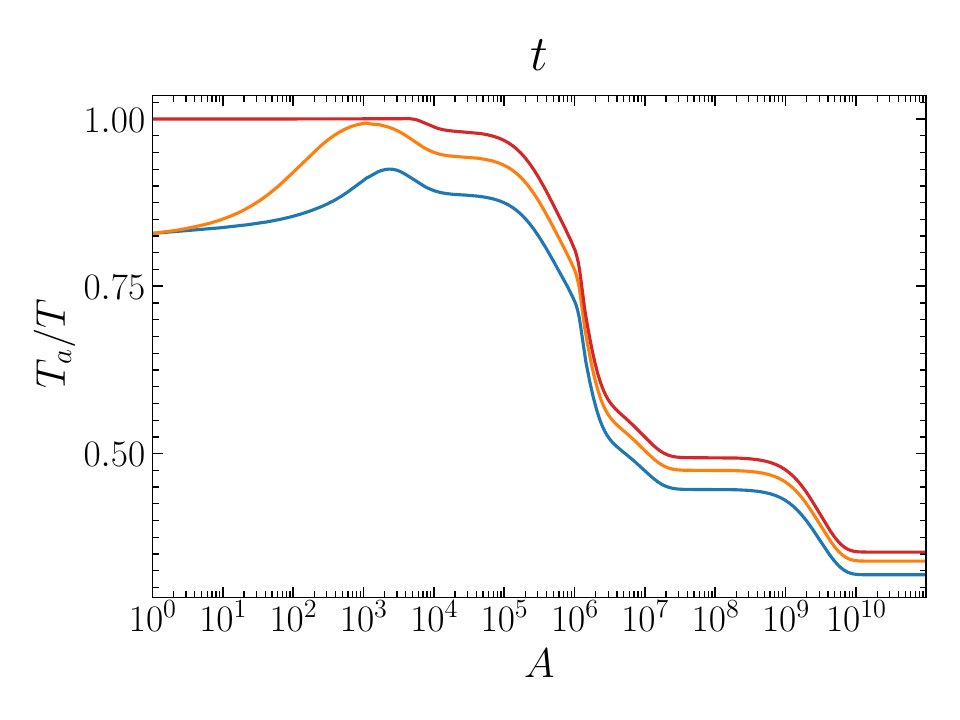}\\
	\includegraphics[width=0.4\textwidth]{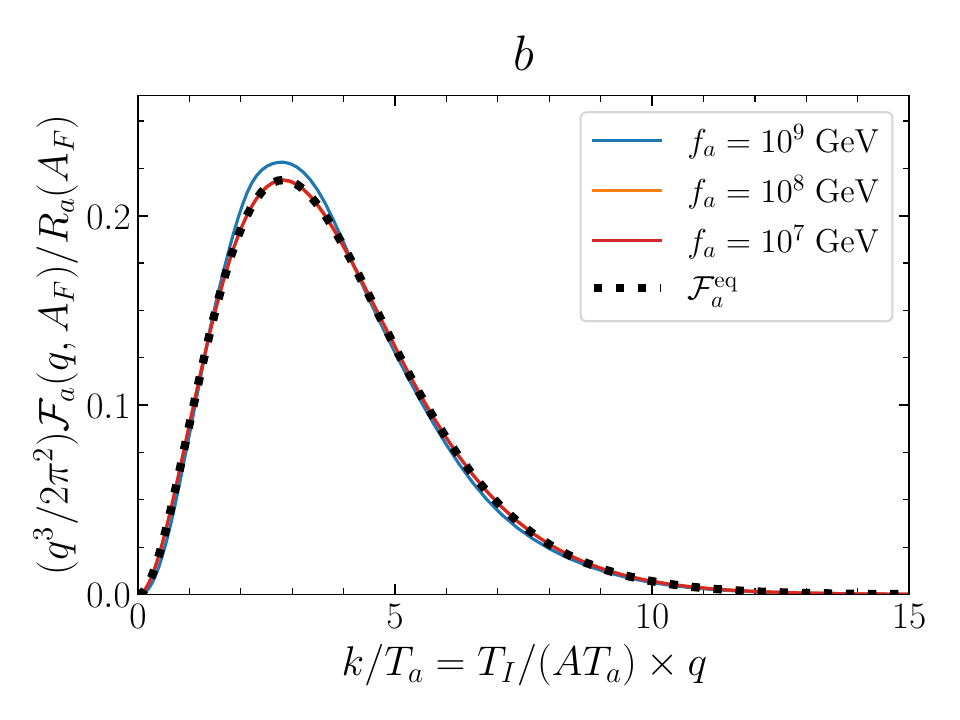} $\qquad$
	\includegraphics[width=0.4\textwidth]{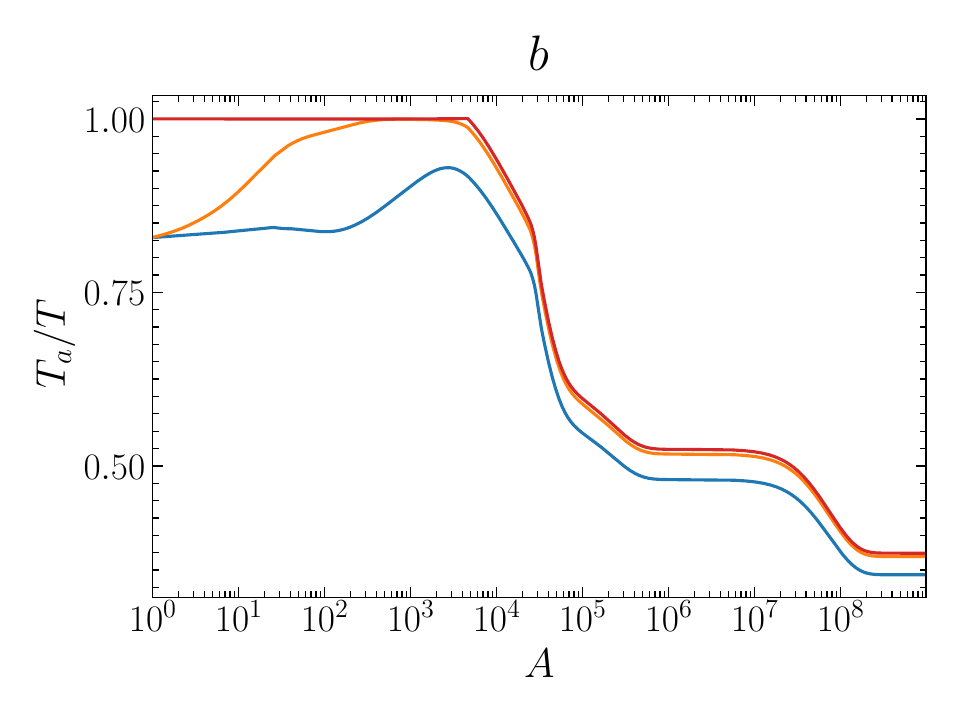}\\
	\includegraphics[width=0.4\textwidth]{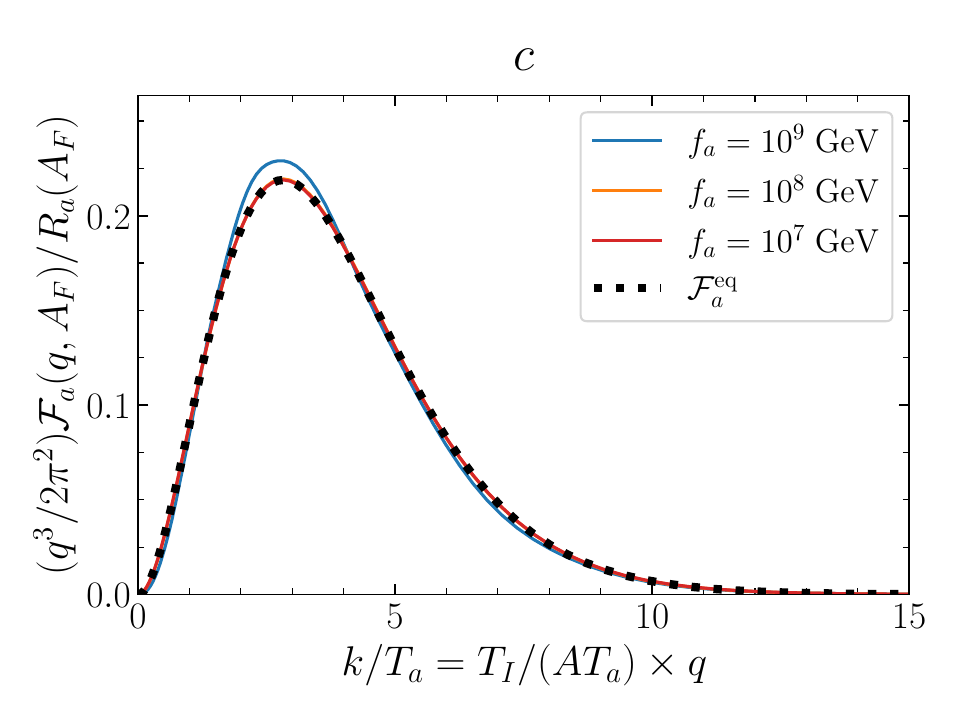} $\qquad$
	\includegraphics[width=0.4\textwidth]{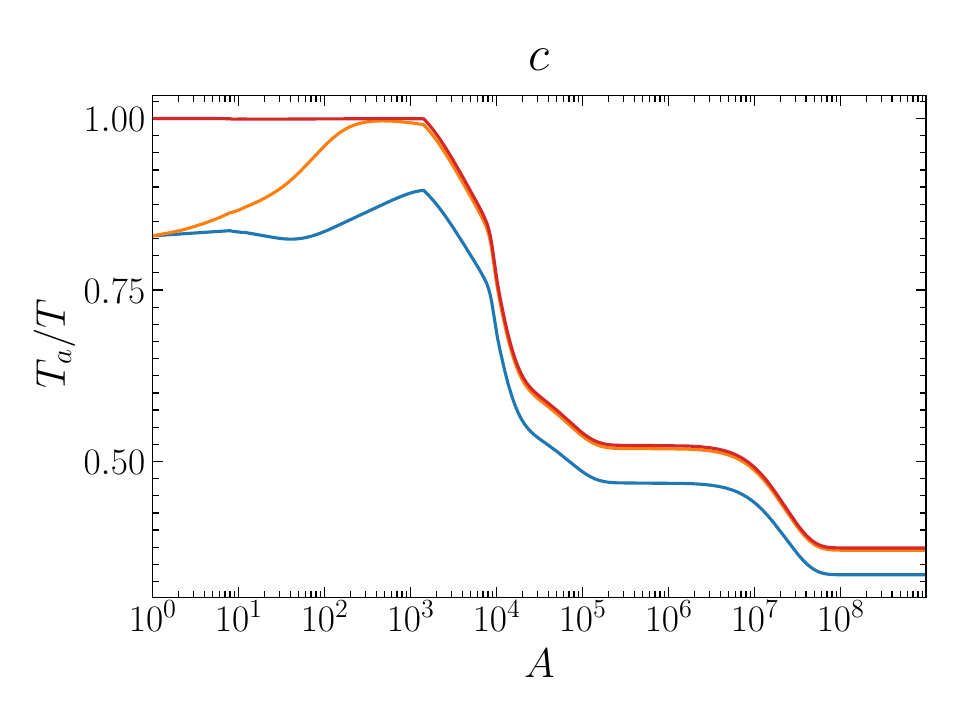}
\caption{Numerical results for hadrophilic axions. Notation as in Fig.~\ref{fig:PSDleptons}.}
\label{fig:PSDquarks}
\end{figure*}

The quark couplings case is structurally similar to the previous one since axion production is driven by the same two classes of channels: quark/antiquark pair annihilations, $q {\overline q} \to g \ax$, and Compton-like scatterings $q g \to q \ax,\ {\overline q} g \to {\overline q} \ax $. The matrix elements calculations are similar with the important addition of color factors. First, we have to multiply them by the combination ${\cal T}_F(N_c^2-1)$. Here, ${\cal T}_F = 1/2$ is the Dynkin index of the fundamental representation of the color group $SU(3)_C$ and $N_c = 3$ is the number of colors. Then we have to account for color degrees of freedom in the average, and this is achieved by dividing the squared matrix elements by an overall factor of $N_c^2 (N_c^2-1) = 72$. We find 
\bea 
|{\cal M}_{q {\overline q} \to g \ax}|^2 &=& \frac{c_q^2 g_s^2}{36f_\ax^2} \frac{m_q^2 s^2}{(m_q^2 - t)(s+t - m_q^2)}\ ,\\
|{\cal M}_{q g \to q \ax}|^2 &=& \frac{c_q^2 g_s^2}{36f_\ax^2}  \frac{m_q^2 t^2}{(s-m_q^2)(s+t - m_q^2)}\ .
\eea
It is worth noticing that the quantities entering the rate calculations are the combinations $g_q^2g_g|{\cal M}_q|^2$, where $g_i$ is the number of internal degrees of freedom of the bath particle $\bath_i$. Thus the production rates for hadrophilic axions can be directly obtained by multiplying the leptophilic results (proportional to the combinations $g_\ell^2g_\gamma|{\cal M}_\ell|^2$) by an overall factor of $4$ and replacing $m_\ell \to m_q$.

The numerical analysis is analogous to the letophilic axion case with a few important caveats. For bottom and charm quarks, there are additional complications due to the vicinity of the confinement scale to the point where production is most efficient and the consequent non-perturbativity of QCD. We are studying the axion production via interactions with quarks, and we have to ensure that these particles are in the thermal plasma. However, as we track the momentum distribution of axions from higher to lower temperatures, we approach the QCD crossover where a drastic change in the degrees of freedom occurs and quarks are confined within hadrons. For this reason we shut off all the interactions of the axion with quarks below the temperature $T_{\rm STOP}= 1\ {\rm GeV}$ by setting all the matrix elements to zero. We argue that this is a way to get a meaningful lower bound on the energy density of axions produced via processes involving the bottom and the charm quark. More details are presented in Sec.~\ref{sec:QCDstop}. Another subtlety is the running of the strong gauge couplings and quark masses. In our calculations, we set $\alpha_s$ and the quark masses $m_q$ at their value at the $\overline{\rm MS}$ renormalization scale $\mu = m_q$. This is reasonable since the IR-production is dominated at temperatures around the quark mass, and the details of the QCD running parameters at lower temperatures $T<m_q$ will be subdominant as most of the axions would have already been produced by the time the plasma has cooled below $T=m_q$. We discuss the uncertainties associated with this procedure in Sec.~\ref{sec:running}.

Fig.~\ref{fig:PSDquarks} shows the axion asymptotic phase space distributions (left panels) and the evolution of the axion temperature (right panels). Not surprisingly, our findings are analogous to the ones in the previous sections. The resulting distributions are very close to a thermal shape even if equilibrium is never achieved, and this is due to the fact that their origin is from the thermal bath. The temperature evolution is also similar, and larger values of the axion decay constant lead to a cooler axion population. The resulting effects in the prediction of $\dneff$ can be found in the next section. We conclude this one with a deeper discussion of the theoretical uncertainties mentioned in the previous paragraph for the bottom and charm quarks. 

\subsection{Approaching the QCD crossover} 
\label{sec:QCDstop}

We investigate the impact of our choice for the temperature $T_{\rm STOP}$ at which we shut off axion interactions. Qualitatively, if this temperature is not hierarchically smaller than the quark mass, the final amount of produced dark radiation will depend on $T_{\rm STOP}$, as axion production ceases at this point. This is relevant for the charm and bottom quarks since the minimum $T_{\rm STOP}$ we allow, 1 GeV (due to confinement), is close to their masses. In contrast, there is no significant effect on the $\dneff$ calculation for the top quark since $m_t \gg T_{\rm STOP}$.

\begin{figure*}[!t]
	\centering
\includegraphics[width=0.44\textwidth]{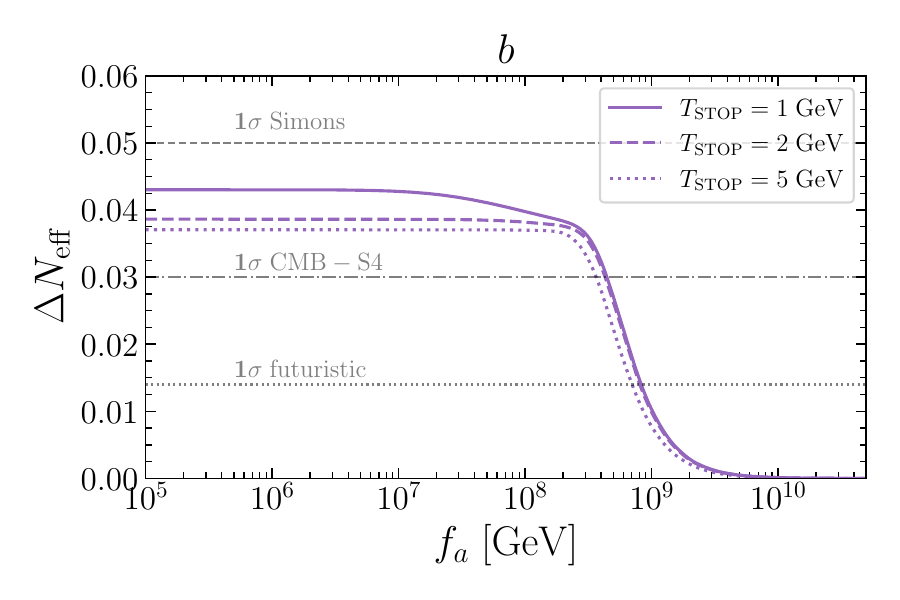} $\qquad$ \includegraphics[width=0.44\textwidth]{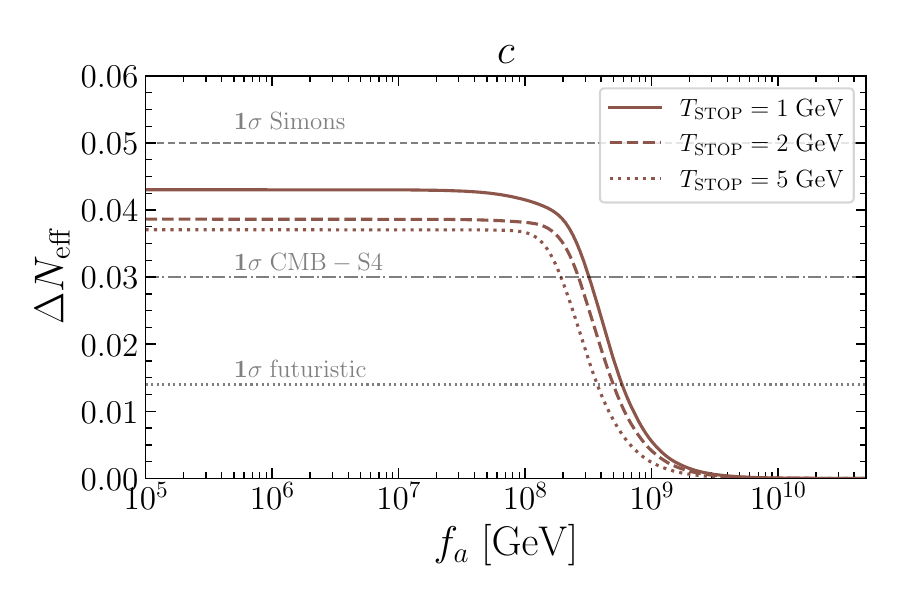}
	\caption{Predicted value of $\dneff$ for different choices of $T_{\rm STOP}=\{1,2,5\}$ GeV, i.e. the temperature at which we shut off the axion-quark coupling. The predictions for the bottom and charm quarks are shown in the left and right panels, respectively.}
	\label{fig:Tstop}
\end{figure*}

The case of the bottom quark is borderline. The mass $m_b \sim 4$ GeV is close, but a factor of few above 1 GeV. For values $f_\ax$ larger than some critical $f_\ax^*$ (i.e. for weaker interactions), most of the momenta decouple at $T=T_{\rm d}>T_{\rm STOP}$ and the amount of produced dark radiation is then only weakly dependent on the actual value of $T_{\rm STOP}$. For stronger couplings, i.e. $f_\ax<f_\ax^*$, the decoupling temperature $T_{\rm d}$ for most of the momenta will be lower than $T_{\rm STOP}$, so the value of $\dneff$ will depend on the actual chosen value of $T_{\rm STOP}$. A good estimate of this behavior is obtained by comparing the decoupling temperature found from $\Gamma(T_{\rm d}) = H(T_{\rm d})$ to $T_{\rm STOP}$ for various values of $f_\ax$. If $T_{\rm d} >T_{\rm STOP}$, the calculation of $\dneff$ can be considered reliable. Otherwise, the obtained $\dneff$ can be only a lower (conservative) bound on the actual value, because $\dneff$ would depend directly on the chosen $T_{\rm STOP}$. The left panel of Fig.~\ref{fig:Tstop} shows $\dneff$ when varying $T_{\rm STOP}$ for the choices $T_{\rm STOP}= \{1,2,5\}$ GeV. We see that the value of $T_{\rm STOP}$ impacts the value of $\dneff$. However, we note that the dependence is most dramatic for $f_\ax<f_\ax^* \sim 4\times 10^8$ GeV. This is because for a larger $f_\ax$, most of the momenta decouple well before the plasma cools down to $T_{\rm STOP}$ or they do not couple at all. For the highest value of $T_{\rm STOP}=5$ GeV, the $\dneff$ is underestimated, as this temperature is larger than the bottom quark mass. 

\begin{figure*}[!t]
	\centering
		\includegraphics[width=0.44\textwidth]{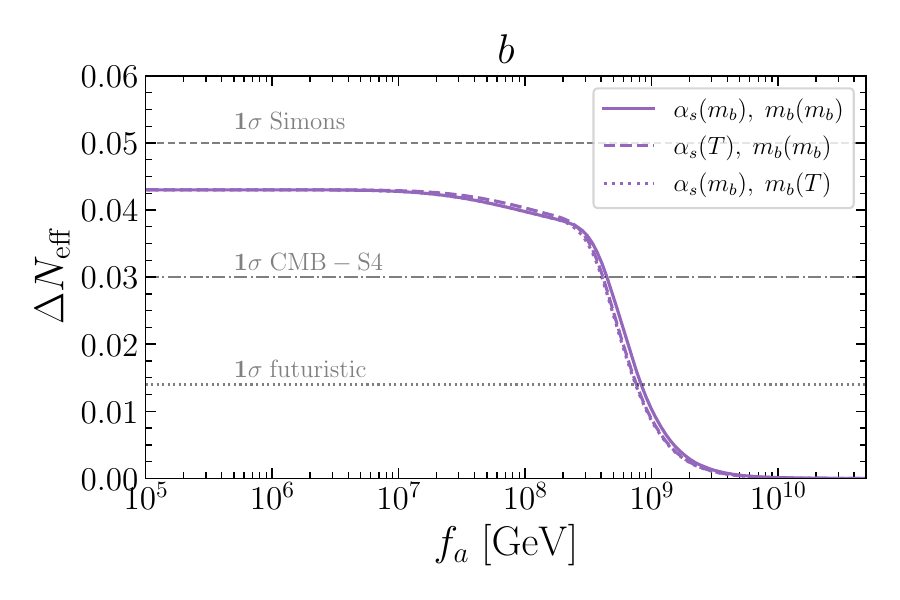} $\qquad$
	\includegraphics[width=0.44\textwidth]{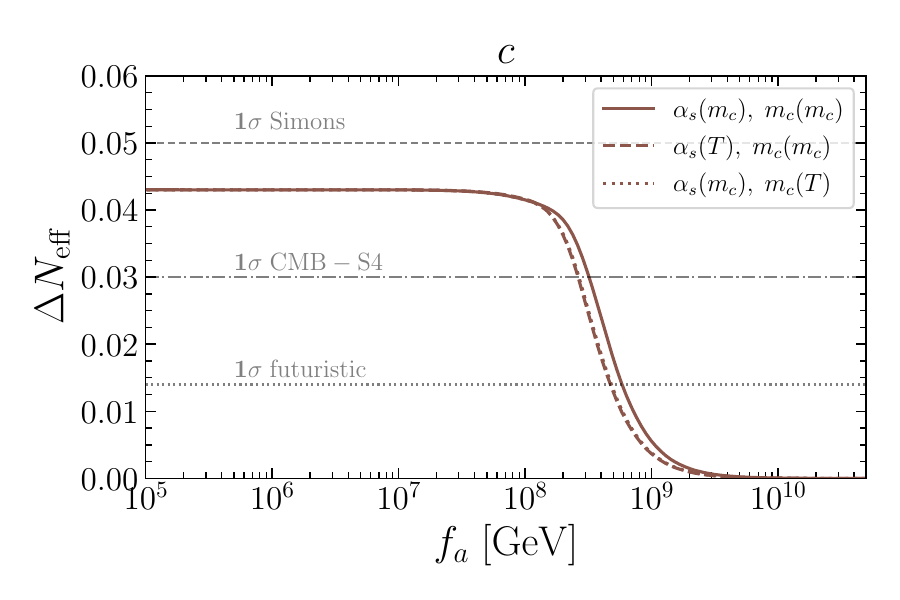}
	\caption{Calculation of $\dneff$ with different running of QCD parameters for bottom (left) and charm (right) quarks. We compare the results presented in this paper (solid lines) where we fix the strong gauge coupling and quark mass with cases where we let run either $\alpha_s(T)$ (dashed) or the quark mass $m_q(T)$ (dotted). }
	\label{fig:running}
\end{figure*}

The charm quark presents the same issue but on a more worrisome level. Being $m_c \simeq 1.3$ GeV, i.e. very close to the lowest value of $T_{\rm STOP}$, it is not possible to find a value of $f_\ax$ for which the momenta decouple well before $T_{\rm STOP}$. For this reason, $T_{\rm STOP}$ acts as an effective instantaneous decoupling temperature enforced artificially before any Boltzmann suppression, and $\dneff$ is a direct function of $T_{\rm STOP}$. For the charm, only a lower bound on $\dneff$ can be computed as illustrated by the right panel of Fig.~\ref{fig:Tstop}. Unlike the bottom case, the computed $\dneff$ are farther away from each other even at large values of $f_\ax$. In particular, to support that $\dneff$ is mostly determined by the value of $T_{\rm STOP}$, the results for small $f_\ax< 10^{7}$ GeV coincide for charm and bottom.

\subsection{QCD running effects}
\label{sec:running}

As the thermal plasma cools down, QCD gets non-perturbative and confinement occurs. We investigate the effects of the evolution of the strong coupling fine structure constant $\alpha_s(T)$ and the quark masses $m_q(T)$, and we consider the running as a function of temperature as every interaction is thermally averaged via bath particle distributions; momenta larger than the temperature are always exponentially suppressed. In our results, we set these running parameters at their values at the mass of the quark under consideration. The rationale is that most of the axion population gets produced before the temperature of the plasma drops below the mass of the quark. This makes our evaluation of $\dneff$ weakly dependent on the couplings and masses at temperatures $T<m_q$ where they would run to higher values. 

To check the goodness of our approximation, we consider the running $\alpha_s(T)$ across the three quark thresholds for the $t,\ b,\ c$ and quark masses $m_q(T)$. We evaluate these running quantities in the $\overline{\rm MS}$ scheme with the software \ttt{CRunDec} \cite{Herren:2017osy} and we use them to compute the collision term and evolve the axion momentum distribution. We focus on the charm and bottom since their masses are very close to the QCD crossover temperature and the effects due to the running are enhanced. We show the results for $\dneff$ with running QCD quantities in Fig.~\ref{fig:running}. We compare our results (solid line) with the other two cases where we run separately only the strong gauge coupling constant and fix the masses to their values at $\mu=m_q$ or, viceversa, we fix $\alpha_s$ and run the quark mass. For both the bottom quark and the charm, the lines lie on top of each other for most values of $f_\ax$, justifying our approximation. The outcome for the charm quark is more sensitive to running effects. This is expected due to the lower mass of the charm quark.

%%%%%%%%%%%%%%%%%%%%%%%%%%%%%%%%%%%%%%%%%%%%%%%% 
\section{Phase Space Impact on $\dneff$}
\label{sec:dneff}
%%%%%%%%%%%%%%%%%%%%%%%%%%%%%%%%%%%%%%%%%%%%%%%% 

 \begin{figure*}[!t]
 	\centering
 	 \includegraphics[width=0.45\textwidth]{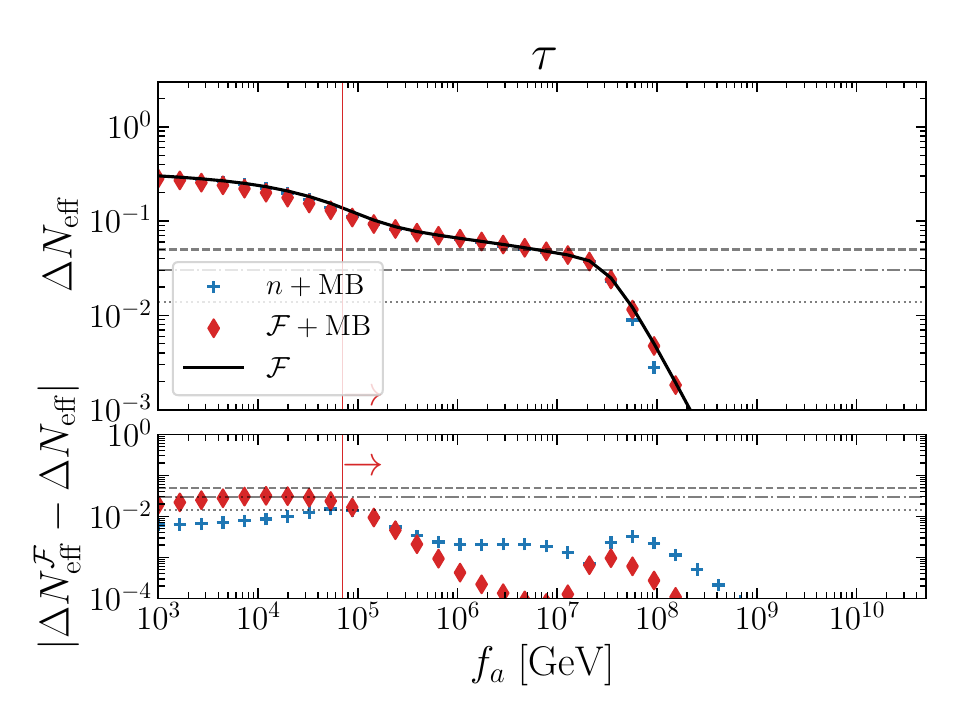} $\qquad$
 	\includegraphics[width=0.45\textwidth]{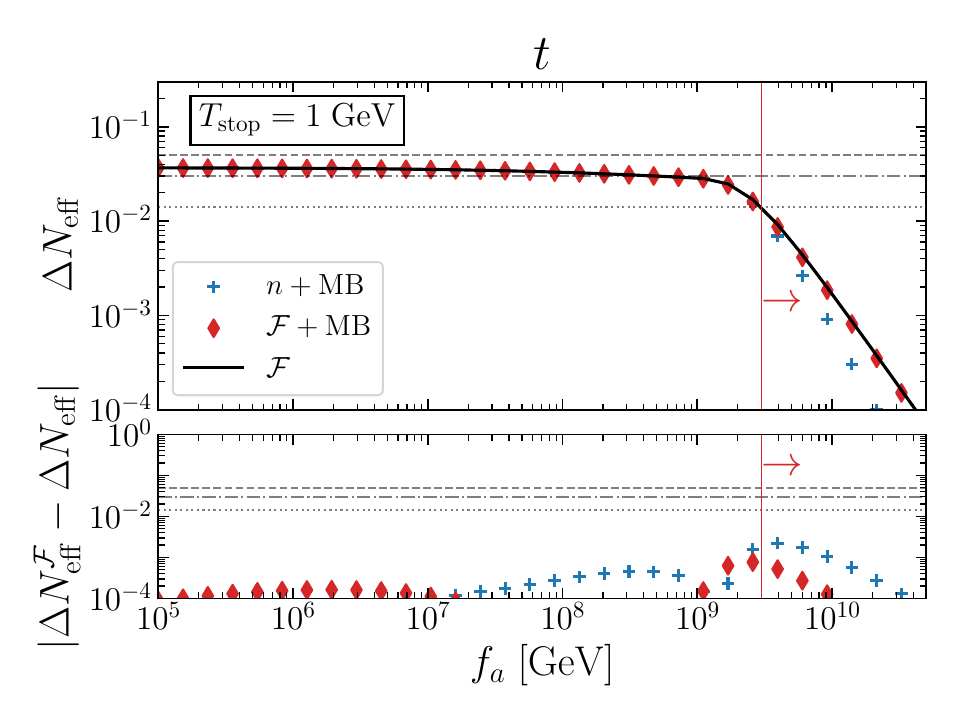}\\ 
 	 \includegraphics[width=0.45\textwidth]{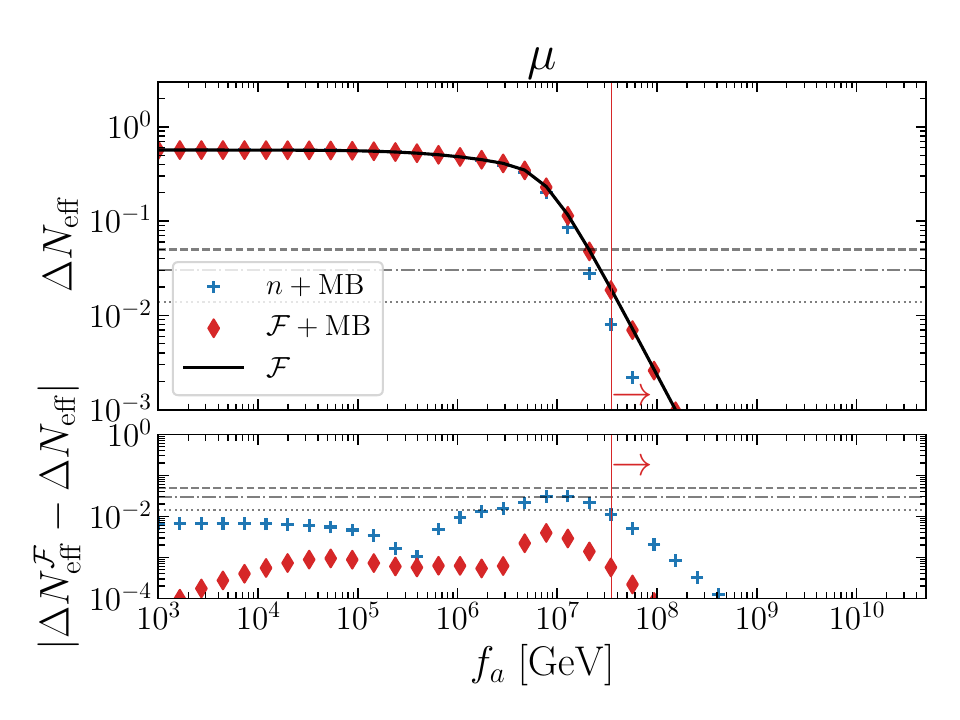}  $\qquad$
 	 	\includegraphics[width=0.45\textwidth]{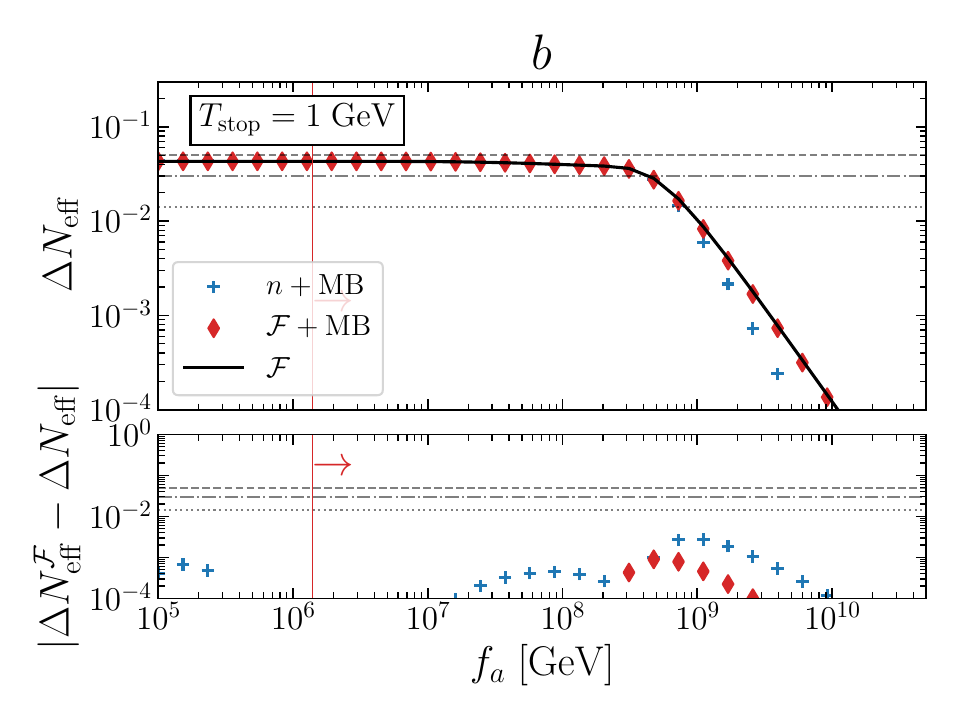}\\ 
 	 \includegraphics[width=0.45\textwidth]{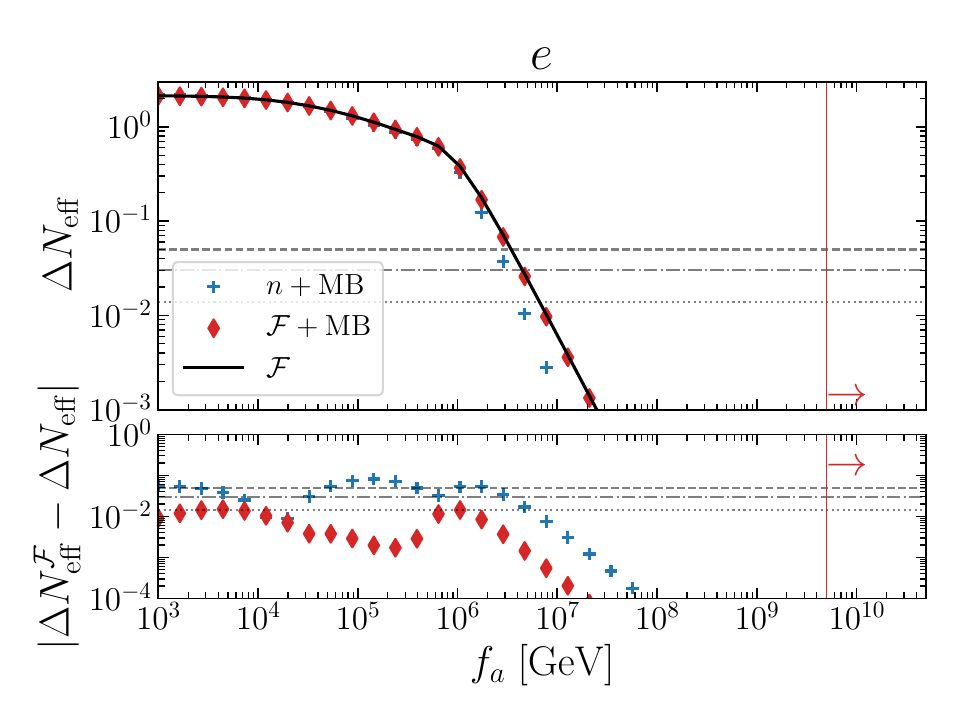}  $\qquad$
\includegraphics[width=0.45\textwidth]{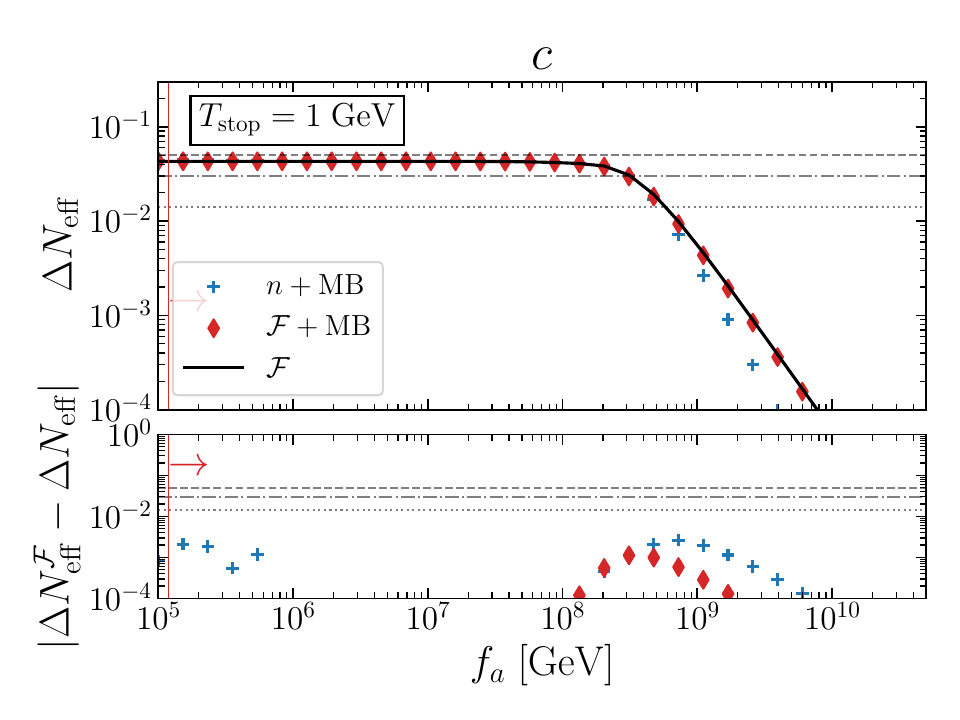}
 	 	 	 	\caption{Summary of the predicted values of $\dneff$ for the leptophilic axion (left column) and the hadrophilic axion (right column). We compare the full phase space solution (black solid lines) with the results obtained in phase space but neglecting the quantum statistics of the thermal bath (red diamonds) and the estimate obtained by tracking the number density (blue crosses). The horizontal lines show the sensitivities, at $1\sigma$ of Simons Observatory, CMB-S4 and CMB-HD. The vertical red line, with arrow, indicates the range of axion decay constants $f_\ax$ allowed by astrophysical constraints (details about the origin of these bounds can be found in App.~\ref{app:bounds}). As manifestly stated in the legenda for the hadrophilic cases, we compute the value of $\dneff$ by shutting off the production rate due to scatterings with quarks and gluons at temperatures below $T_{\rm STOP}=1$ GeV. The lower panel in each plot shows the absolute error between the phase space solution and the two employed approximations.}
\label{fig:dneffcomparison}
 \end{figure*}

The momentum distributions determined in the two previous sections for leptophilic and hadrophilic axions are the essential input to determine $\dneff$ via the expression in Eq.~\eqref{eq:dneffinal}. The factor $R_\ax(A_F)$ is given explicitly by Eq.~\eqref{eq:Rax} as an integral over the asymptotic phase space distribution for the comoving momentum. In this section, we report the predicted values of $\dneff$ as a function of the axion decay constant $f_\ax$ for the six different cases discussed above. Our interest here is not only in the overall size of the effect, but also whether the results obtained via this more elaborated machinery represent a meaningful improvement compared to previous studies. This is why we predict $\dneff$ also via two different kind of approximate analysis and compare these results with our findings. We refer to Sec.~\ref{sec:leptophilic} and  \ref{sec:hadrophilic} for the details of the calculations and, in the hadrophilic case, the caveats and the assumptions related to the QCD crossover.

The six panels in Fig.~\ref{fig:dneffcomparison} show $\dneff$ as a function of the axion decay constant. In addition to the results arising from a complete analysis (labeled as $\cal F$), we also show predictions based on the standard solution of the Boltzmann equation for the number density (${n}+{\rm MB}$) -- a method review by App.~\ref{app:BE4n} -- and via a phase space analysis that neglects quantum effects of the thermal bath (${\cal F}+{\rm MB}$). The $\dneff$ for the hadrophilic axion is obtained by shutting off the interaction between axions and quarks and gluons below $T_{\rm STOP}=1$ GeV (see Sec~\ref{sec:QCDstop} for a discussion). For each fermion we also present a subpanel with the absolute error between the full phase space solution and the two approximate methods. The vertical red lines show the allowed parameter space in $f_\ax$ according to the astrophysical constraints collected in App.~\ref{app:bounds}. The horizontal grey lines mark the 1$\sigma$ sensitivities of Simons Observatory \cite{SimonsObservatory:2018koc}, CMB-S4 \cite{CMB-S4:2016ple,Abazajian:2019eic,CMB-S4:2022ght} and futuristic proposals~\cite{Sehgal:2020yja}; these lines serve as a guide to establish whether the differences between the methods to compute $\dneff$ are meaningful.

The leptophilic case is particularly intriguing. The $\dneff$ signal lies in the observable region (i.e. above the grey lines) in the allowed parameter space (on the right of the red line) for the $\tau$ and the $\mu$. Moreover, the differences between methods indicate that the full phase space solution is essential to reach reliable predictions for future observations. The hadrophilic axion scenario has also interesting messages. The parameter space for the top quark is in tension with astrophysical bounds, and the $\dneff$ for the allowed decay constants lies at the edge of the sensitivity reach of future CMB observatories. While this is not the case for the bottom and the charm quark, these two particles suffer more for the uncertainties related to QCD. In any case, it seems that different methods produce values for $\dneff$ that differ by an unobservable amount.

\begin{figure}[t!]\includegraphics[width=0.46\textwidth]{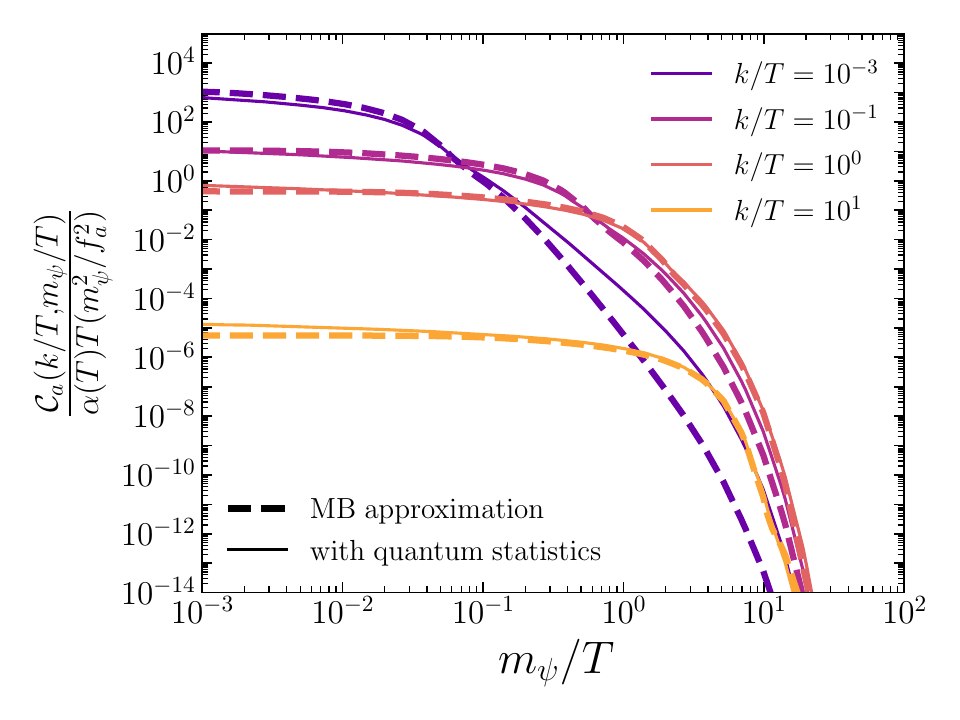}\caption{The (normalized) total collision term for axion production through binary scatterings. We show the collision term for different comoving momenta $k/T$ comparing the result using the MB approximation (thick dashed lines) and the full numerical result including the quantum effects. This plot is valid for leptons, the analogous one for quarks would be identical up to an overall factor of $4$.  }\label{fig:statistics}	
\end{figure}

Another important point can be evinced from Fig.~\ref{fig:dneffcomparison}: the impact of quantum statistics of the bath is minimal and in almost all cases, unobservable. The $\tau$ lepton is an exception: it has a mass that lies very close the QCD crossover where the number of relativistic degrees of freedom rapidly changes, amplifying differences in the collision terms due to e.g. to quantum statistics. The crucial quantity is the temperature of decoupling of each momentum bin, set by the relative size of the collision term ${\cal C}_\ax(k,T)$ with respect to the Hubble rate. In  Fig.~\ref{fig:statistics}, we show the normalized collision term for the processes we will consider, namely binary scatterings $\psi \gamma \to \psi a$ and $\obar \psi \gamma \to \obar\psi a$ (annihilations $\psi \obar\psi \to \gamma a$) of fermions with
(into) gauge bosons, producing axions. We compare the full Maxwell-Boltzmann approximation, that neglects the statistics of the bath particles with the full numerical result for the quantum statistics considered in this work. We see that ${\cal C}_\ax(k,T)$ is only very slightly affected by the quantum statistics. As shown already in Ref.~\cite{DEramo:2023nzt}, this is the case unless $g_{*\rho}(T)$ varies rapidly around the peak of axion production, $T \sim m_\psi$: in such cases a difference in the collision term could result in a delayed decoupling of some comoving momenta and in spectral distorsions associated with variations in $\dneff$.

\section{Conclusions}
\label{sec:conclusions}

\begin{figure*}[!t]
\centering
\includegraphics[width=0.49 \textwidth]{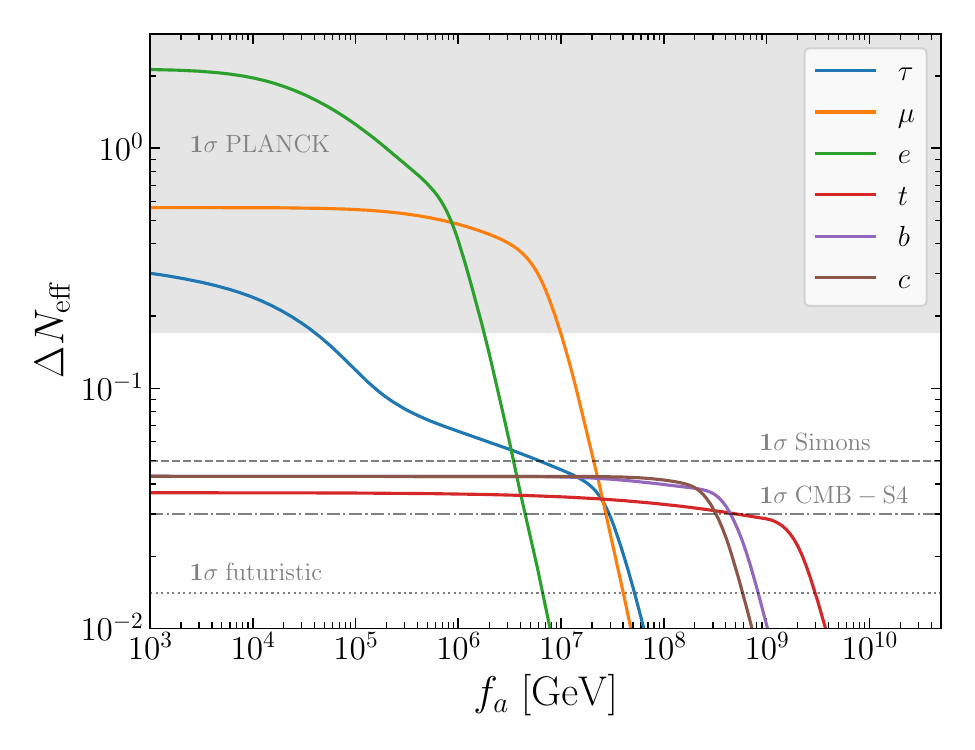} $\;\,$
\includegraphics[width=0.49\textwidth]{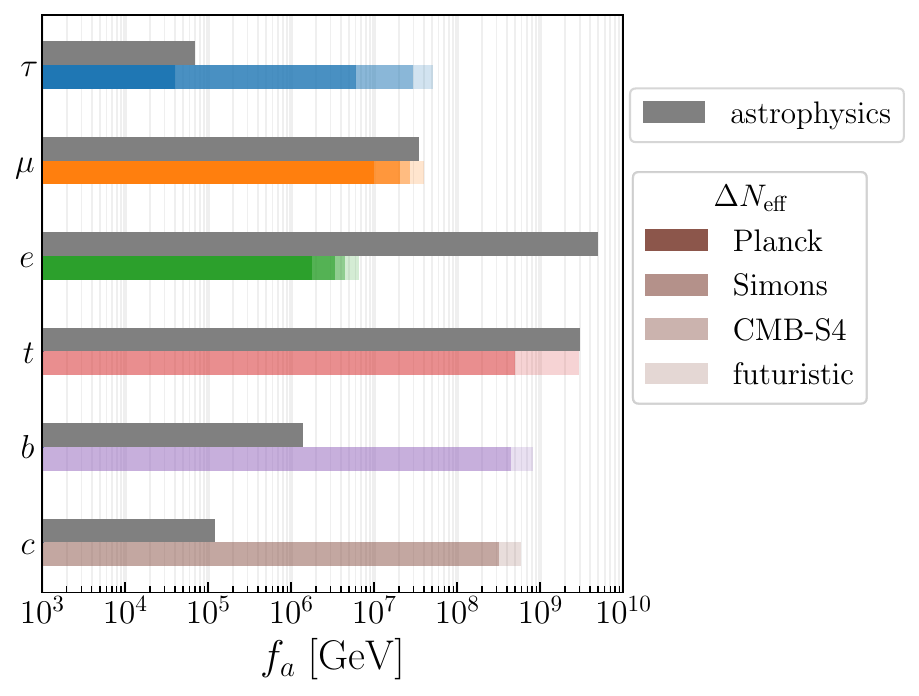}
\caption{Summary of the main results achieved in this work. The left panel shows the predicted $\dneff$ through a full phase space analysis as a function of $f_\ax$ for the three charged leptons and the three heavy quarks. It is understood that only one dimensionless Wilson coefficient is switched on at a time and set to $c_\psi = 1$. The right panel shows current and future cosmological constraints on the axion decay constant obtained through a full phase space analysis. As in Fig.~\ref{fig:NeffFinal}, only one Wilson coefficient is switched on and set to $c_\psi = 1$. Different color shadings show the Planck, Simons, CMB-S4, and futuristic constraints. For comparison, we show with gray bands also the bounds from astrophysics.}
\label{fig:NeffFinal}
\end{figure*}

We investigated the cosmological consequences of a light pseudo-scalar $\ax$ coupled to standard model fermions via higher-dimensional operators. Our conceptual starting point was a UV complete theory where the axion arises as a Nambu-Goldstone boson from the spontaneous breaking of a global Abelian symmetry. For the specific case of the PQ symmetry we have the QCD axion solving the strong CP problem. We were agnostic in our study about the UV origin since at low enough energies the axion is the only leftover new degree of freedom. As we discussed quantitatively in Sec.~\ref{sec:thermalization}, axion interactions with fermions lead to a IR-dominated thermalization and for this reason the focused our study on the effective interactions in Eq.~\eqref{eq:LintPsi} that are left once we focus on physics at energies below the weak scale.

The main output of this work is the precise calculation of the number of effective neutrino species,  $\dneff$, defined in Eq.~\eqref{eq:dneff} in presence of a QCD axion with interactions with SM fermions. We considered the cases of leptophilic and hadrophilic axions, that is interactions with charged leptons $\{\tau,\mu,e\}$ and heavy quarks $\{t,b,c\}$, respectively.

We summarize the key findings in Fig.~\ref{fig:NeffFinal}. In the left panel we present the predicted $\dneff$ obtained through a full phase space calculation and accounting for quantum statistics. The case of axions produced via processes involving quarks is delicate, and we refer to Sec.~\ref{sec:hadrophilic} for a discussion on the details of the computation. We overline present and future $1\sigma$ constraints on $\dneff$ from observations of the Cosmic Microwave Background (CMB). The gray band indicates the region already excluded by Planck \cite{Planck:2018vyg}, while the horizontal lines show the reach sensitivities of the Simons Observatory \cite{SimonsObservatory:2018koc}, CMB-S4 \cite{CMB-S4:2016ple,CMB-S4:2022ght} and futuristic proposals~\cite{Sehgal:2020yja}. While Planck already excludes a significant part of the parameter space in $f_\ax$ for leptons, heavier quarks are beyond its reach and can be probed only by future observations.

We visualize the global status of constraints on the axion decay constant (via couplings to the fermions) in the right panel of Fig.~\ref{fig:NeffFinal}. The gray bands show constraints from astrophysics, that are very effective in excluding a large part of parameter space for electrons, top quarks and muons couplings to axions. Via such a direct comparison, we see that CMB observations can provide competitive (with astrophysics) bounds on $f_\ax$ for most fermions. The case of the $\tau$ lepton is particularly intriguing as the Planck constraint is already on par with the astrophysical bound and can be improved by more than two orders of magnitude with future CMB observations. The bottom and charm quarks also deserve particular attention, as CMB is still not able to constraint their couplings to axions. We argued in Sec.~\ref{sec:QCDstop} that $\dneff$ obtained for bottom and charm quarks can just be thought in most cases as a lower bound on the axion energy density present at the CMB formation. Nevertheless, these results can be used effectively to put very strong and conservative bounds on the axion-quark derivative operator once future CMB data become available.

Light and weakly-coupled PNGB remain among the most appealing candidates for physics beyond the standard model. The QCD axion, which solves the strong CP problem and provides a viable cold dark matter candidate, is a prominent example. The primordial thermal bath provides a powerful and complementary probe for axion couplings to standard model matter, and the next decades will witness significant improvements on the sensitivity on $\dneff$. Our analysis supports the need for rigorous theoretical phase space analysis to match this astonishing experimental effort.

\vspace{0.2cm}

\textbf{NOTE ADDED.} While we were finalizing this work, Ref.~\cite{Badziak:2024qjg} appeared on the arXiv which performed a phase space analysis of flavor-diagonal and flavor-violating axion couplings to leptons. 

\vspace{-0.4cm}

\acknowledgments
The authors thank Miguel Escudero, Rotem Ovadia and Luca Vecchi for useful discussions. We also thank the journal referee for their constructive feedback on the first version of this manuscript. This work is supported in part by the Italian MUR Departments of Excellence grant 2023-2027 ``Quantum Frontiers''. F.D. is supported by Istituto Nazionale di Fisica Nucleare (INFN) through the Theoretical Astroparticle Physics (TAsP) project. F.D. acknowledges support from the European Union’s Horizon 2020 research and innovation programme under the Marie Skłodowska-Curie grant agreement No 860881-HIDDeN. The work of A.L. is supported by an ERC STG grant (``Light-Dark'', grant No. 101040019). 
This project has received funding from the European Research Council (ERC) under the European Union’s Horizon Europe research and innovation programme (grant agreement No. 101040019).  Views and opinions expressed are however those of the author(s) only and do not necessarily reflect those of the European Union. The European Union cannot be held responsible for them. A.L. is also in part supported by the US-Israeli BSF grant 2016153. A.L. is grateful to the CERN theory group for the support while this work was carried out.

\appendix 

%%%%%%%%%%%%%%%%%%%%%%%%%%%%%%%%%%%%%%%%%%%%%%%% 
\section{Bounds on Axion-Fermion Couplings}
\label{app:bounds}
%%%%%%%%%%%%%%%%%%%%%%%%%%%%%%%%%%%%%%%%%%%%%%%% 

The strongest bounds on flavor conserving couplings to standard model fermions come from astrophysics. For the two lightest leptons, the bounds come from observations of astrophysical environments where $e$ and $\mu$ play an active role. Brightness calibrations of the tip of the red-giant branch and the $\omega$-Centauri red-giant give $f_\ax/c_e \gtrsim 5 \times 10^9\  {\rm GeV}$~\cite{Capozzi:2020cbu}. Energy loss in SN 1987A~\cite{Caputo:2021rux} and HB stars cooling in globular clusters~\cite{Carenza:2020zil} lead to $f_\ax/c_\mu \gtrsim 3.5 \times 10^7\ {\rm GeV}$. For the other fermions, the strongest bounds still come from astrophysics as showed by Ref.~\cite{Feng:1997tn} via a one-loop induced coupling to electrons. We follow the notation and calculations of App.~B of Ref.~\cite{DEramo:2018vss}. The Wilson coefficient $c_\psi$ appearing in Eq.~\eqref{eq:LintPsi} is generated at some UV cutoff scale $\Lambda\sim f_\ax$, and the consequent low-energy electron coupling results in
\bea 
c_e = c_e(\Lambda) -\frac{y_\psi^2}{8\pi^2}N_c^\psi c_\psi\log(\Lambda/m_\psi) \ ,
\eea 
where $y_\psi$ and $N_c^\psi$ are the fermion Yukawa coupling and its number of colors, respectively. The most conservative bound is obtained by assuming that the axion does not couple to electrons at the UV scale (i.e., $c_e(\Lambda)=0$). We set $\Lambda = f_\ax$ and find the  constraints 
\be
f_\ax / c_\psi \gtrsim {\rm GeV} \times \left\{ \begin{array}{lccc}
3.1 \times 10^9\  & & & \psi = t \\
1.4 \times 10^6 & & & \psi = b \\
1.2 \times 10^5 & & & \psi = c \\
7.0 \times 10^4 & & & \psi = \tau
\end{array}
\right. \ .
\ee
Regarding muons, the same approach would bring a non competitive value, suppressed by a factor $m_\mu^2/m_\tau^2 \sim 100$ with respect to the tau one.

%%%%%%%%%%%%%%%%%%%%%%%%%%%%%%%%%%%%%%%%%%%%%%%% 
\section{Tracking the axion number density}
\label{app:BE4n}
%%%%%%%%%%%%%%%%%%%%%%%%%%%%%%%%%%%%%%%%%%%%%%%% 

We review the most commonly employed approximate method that tracks the axion number density
\be
n_\ax(t) \equiv g_\ax \int \frac{d^3 k}{(2 \pi)^3} \f_\ax(k, t) \ .
\label{eq:naxdef}
\ee

Ref.~\cite{DEramo:2023nzt} spelled out the details of the derivation from the Boltzmann equation in momentum space for $\f_\ax(k, t)$ as well as the needed approximations. One has to assume both kinetic equilibrium and the Maxwell-Boltzmann statistics for the axion to find
\be
\frac{d n_\ax}{d t} + 3 H n_\ax = \gamma_\ax \left( 1 - \frac{n_\ax}{n_\ax^{\rm eq}} \right) \ .
\label{eq:BEn}
\ee
The dimension four quantity $\gamma_\ax$ describes the number of axions produced per unit time and volume. Consistently with the Maxwell-Boltzmann statistics for the axion, an assumption needed to achieve Eq.~\eqref{eq:BEn}, we have for a generic two-body collision $\bath_i \bath_j \to \bath_\ell \ax$ the expression

\bea\nonumber
\gamma_{\bath_i \bath_j \to \bath_\ell \ax} & = & \int d\Pi_id\Pi_j d\Pi_\ell d\Pi_\ax \\ && \times (2\pi)^4 \delta^{(4)}(P_i + P_j  -P_\ell -K)\\\nonumber
&&  \times |{\cal M}_{\bath_i \bath_j \to \bath_\ell \ax}|^2 \f_{i}\f_{j}(1\pm \f_{\ell}) \ . 
\eea 
Here, we integrate over the Lorentz invariant phase space measures $d \Pi_{i,j,k,a}$ that include the degeneracy factors accounting for the internal degrees of freedom (e.g., spin, colors, etc.). The production process is assumed to be invariant under the time-reversal symmetry T (or, equivalently, CP invariant), and the squared matrix element $|{\cal M}_{\bath_i \bath_j \to \bath_\ell \ax}|^2$ is averaged over both initial and final degrees of freedom as it is conventional in the literature.

Solving the ordinary differential equation in Eq.~\eqref{eq:BEn} is simple and fast, but the explicit evaluation of the production rates is rather expensive unless we make additional approximations. A commonly adopted way to speed up the calculation is to drop the quantum statistical effects also for the bath particles. We employ the Maxwell-Boltzmann statistics for all the particles in the process. We identify the center of mass energy for the process $s \equiv (p_i + p_j)^2$ and define $\lambda(x,y,z) \equiv [x - (y + z)^2]  [x - (y - z)^2]$. We relate the production rate to the Lorentz invariant cross section $\sigma_{\bath_i \bath_j \to \bath_\ell \ax}(s)$ as follows 
\bea\nonumber
\gamma^{\rm MB}_{\bath_i \bath_j \to \bath_\ell \ax} & = & 2 \int d\Pi_id\Pi_j \exp\left[ - (E_i + E_j) / T \right] \\  &&  \times \sqrt{\lambda(s,m_i,m_j)}  \, \sigma_{\bath_i \bath_j \to \bath_\ell \ax}(s)  \ . 
\eea 
Some integrals are straightforward. We employ polar coordinates for both momenta, and the only non-trivial angular integral is the one over the angle $\theta$ between the two initial spatial momenta defined by $s = m_i^2 + m_j^2 + 2 (E_i E_j - p_i p_j \cos\theta)$. Furthermore, the full expression can be simplified through a convenient change of variables. We trade the integrals over $(E_i, E_j, \theta)$ with the ones over the variables $(E_+ \equiv E_i+ E_j, E_- \equiv  E_i-E_j, s)$. After accounting for the Jacobian on this change of variables (see, e.g., App.~A of Ref.~\cite{DEramo:2017ecx}) we find the measure
\be
d\Pi_id\Pi_j  = \frac{g_i  d^3 p_i \, g_j d^3 p_j}{(2 \pi)^6 \, 2 E_i \, 2 E_j} = \frac{g_i g_j dE_+ dE_- ds}{128 \pi^4} \ .
\ee
The integral over $E_-$ is straightforward, the one over $E_+$ produces the Bessel function $K_1$ and the final expression
\bea\nonumber
\gamma_{\bath_i \bath_j \to \bath_\ell \ax}^{\rm MB} &=& \dfrac{g_i g_j T}{32 \pi ^4}\int_{(m_i+m_j)^2}^\infty \frac{ds}{\sqrt{s}} \; K_1[\sqrt{s}/T]  \\ && \times 
\lambda(s,m_i,m_j) \, \sigma_{\bath_i \bath_j \to \bath_\ell \ax}(s) \ .
\label{eq:rateMB} 
\eea

For axion couplings to leptons, the cross sections for the leading production processes read 
\bea 
\sigma_{\ell^+ \ell^-\to \gamma a}(s) &=& \dfrac{c_\ell^2 e^2}{f_\ax^2} \dfrac{x_\ell \, \tanh ^{-1}\left(\sqrt{1-4 x_l}\right)}{4 \pi (1-4 x_l)} \ , \\
\sigma_{\ell^\pm \gamma \to \ell^\pm a}(s) &=& \dfrac{c_\ell^2 e^2}{f_\ax^2} \dfrac{x_\ell (4 x_\ell - 2 x_\ell - x_\ell^2 - 3)}{32 \pi (1 - x_\ell)} .
\eea 
The dimensionless quantity $x_\ell \equiv m^2_\ell / s$ is defined as the ratio between the squared lepton mass $m^2_\ell$ and the Mandelstam variable $s$. For axion couplings to quarks, the analogous processes with gluons are dominant given the hierarchy between the gauge couplings. The expressions for the cross sections are similar, they differ only by color factors and they explicitly read
\bea 
\sigma_{q \overline{q}\to g a}(s) &=& \, \dfrac{ c_q^2 g_s^2}{f_\ax^2} \dfrac{x_q \,\tanh ^{-1}\left(\sqrt{1-4x_q}\right)}{9 \pi (1-4 x_q)} \ , \\
\sigma_{q g \to q a}(s) &=& \, \dfrac{c_q^2 g_s^2}{f_\ax^2}  \dfrac{x_q (4 x_q - 2 \ln x_q - x_q^2 - 3)}{192 \pi (1-x_q)}  \ .
\eea

\begin{figure}[!t]
\centering
	\includegraphics[width=0.46\textwidth]{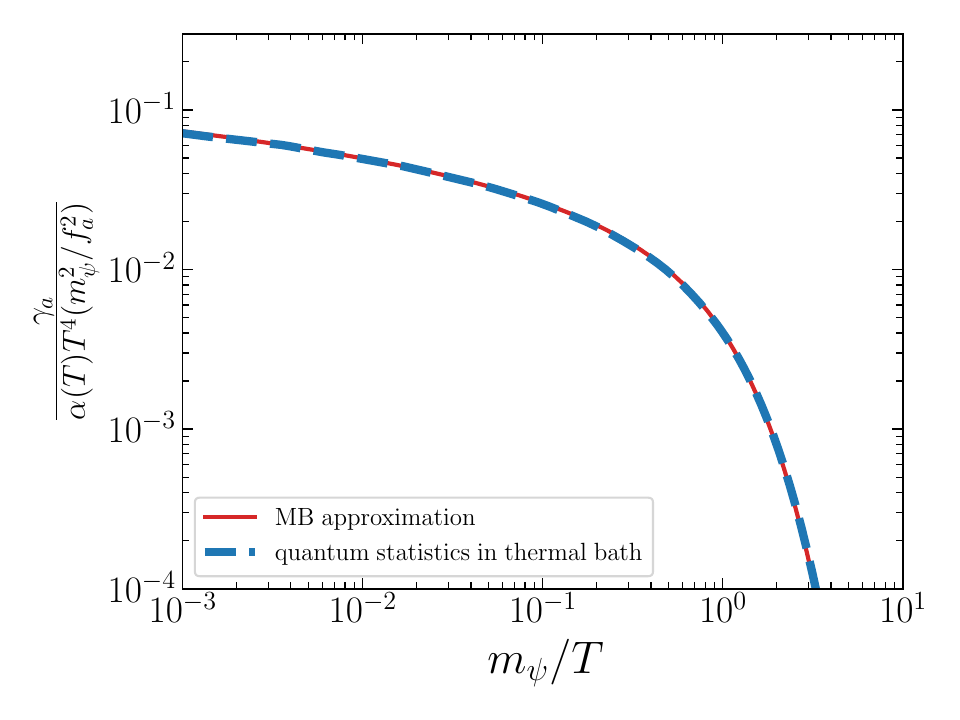}
\caption{Scattering rate $\gamma_a$ appearing in the Boltzmann equation for the number density as a function of the bath temperature $T$. The normalization is chosen to make sure that the results are valid for a generic fermion $\psi$. Solid red solid lines are obtained by treating bath particles with Maxwell-Boltzmann statistics, dot-dashed blue lines include quantum effects. This plot is valid for leptons, the analogous one for quarks would be identical up to an overall factor of $4$.}
\label{fig:statistics_n}
\end{figure}

We conclude this appendix by evaluating the goodness of the Maxwell-Boltzmann approximation for bath particles. As already emphasized, this approximation is unavoidable for the axion particle itself if one wants to get an ordinary differential equation self-consistently. Fig.~\ref{fig:statistics_n} shows the temperature dependence of the total production rate for the two different cases when bath particles are treated classically and quantum mechanically. The convenient normalization is such that these results are valid for any fermion and coupling strengths. The plot is produced for the lepton case, the quark case would differ by only a factor of $4$. As it is manifest from this figure, the Maxwell-Boltzmann statistics is an excellent assumption for bath particles as well.

\vspace{0.2cm}

\bibliography{AxionPSD}
\end{document}